\documentclass[usenatbib,usegraphicx]{mn2e}

\title[Gravitational Lensing by N Point Mass]
{
Perturbation theory of N point-mass gravitational lens systems 
without symmetry: 
small mass-ratio approximation
}
\author[
H. Asada]
{H. Asada
\\ 
Faculty of Science and Technology, 
Hirosaki University, Hirosaki 036-8561, Japan}
\begin{document}

\date{Accepted  Received }

\pagerange{\pageref{firstpage}--\pageref{lastpage}} 

\pubyear{2008}

\maketitle

\label{firstpage}

\begin{abstract}
This paper makes the first systematic attempt to determine  
using perturbation theory 
the positions of images by gravitational lensing due to 
arbitrary number of coplanar masses without any symmetry 
on a plane, as a function of lens and source parameters.  
We present a method of Taylor-series expansion 
to solve the lens equation under a small mass-ratio approximation.  
First, we investigate perturbative structures of 
a single-complex-variable polynomial, 
which has been commonly used. 
Perturbative roots are found. 
Some roots represent positions of lensed images, while the others 
are unphysical because they do not satisfy the lens equation.
This is consistent with a fact that 
the degree of the polynomial, namely the number of zeros, 
exceeds the maximum number of lensed images 
if N=3 (or more). 
The theorem never tells which roots are physical (or unphysical). 
In this paper, unphysical ones are identified. 
Secondly, to avoid unphysical roots, 
we re-examine the lens equation. 
The advantage of our method is that it allows 
a systematic iterative analysis. 
We determine image positions for binary lens systems 
up to the third order in mass ratios 
and for arbitrary N point masses up to the second order. 
This clarifies the dependence on parameters. 
Thirdly, the number of the images that admit a small mass-ratio limit 
is less than the maximum number. 
It is suggested that positions of extra images could not be expressed 
as Maclaurin series in mass ratios. 
Magnifications are finally discussed. 
\end{abstract}


\section{Introduction}
Gravitational lensing has become one of important subjects 
in modern astronomy and cosmology 
(e.g., Schneider 2006, Weinberg 2008). 
It has many applications as gravitational telescopes 
in various fields ranging from extra-solar planets  
to dark matter and dark energy at cosmological scales. 
This paper focuses on gravitational lensing 
due to a N-point mass system.   
Indeed it is a challenging problem to express 
the image positions as functions of lens and source parameters. 
There are several motivations. 
One is that gravitational lensing 
offers a tool of discoveries and measurements 
of planetary systems 
(Schneider and Weiss 1986, Mao and Paczynski 1991, 
Gould and Loeb 1992, Bond et al. 2004, 
Beaulieu et al. 2006),  
compact stars, or a cluster of dark objects, 
which are difficult to probe with other methods. 
Gaudi et al. (2008) have recently found an analogy 
of the Sun-Jupiter-Saturn system by lensing. 
Another motivation is theoretically oriented. 
One may be tempted to pursue a transit between a particle method 
and a fluid (mean field) one. 
For microlensing studies, particle methods are employed, 
because the systems consist of 
stars, planets or MACHOs. 
In cosmological lensing, 
on the other hand, light propagation is considered 
for the gravitational field produced by 
inhomogeneities of cosmic fluids, say galaxies 
or large scale structures of the universe 
(e.g., Refregier 2003 for a review). 
It seems natural, though no explicit proof has been given, 
that observed quantities computed by continuum fluid methods 
will agree with those by discrete particle ones  
in the limit $N \to \infty$, 
at least on average, 
where $N$ is the number of particles. 

Related with the problems mentioned above, 
we should note an astronomically important effect 
caused by the finiteness of $N$. 
For most of cosmological gravitational lenses 
(both of strong and weak ones), 
a continuum approximation can be safe and has worked well.  
There exists an exceptional case, however, 
for which discreteness becomes important. 
One example is a quasar microlensing 
due to a point-like lens object, which is possibly a star 
in a host galaxy (for an extensive review, Wambsganss 2006). 
A galaxy consists of very large number $N$ particles, 
and light rays from an object at cosmological distance 
may have a chance to pass very near one of the point masses. 
As a consequence of finite-$N$ effect 
in large $N$ point lenses, 
anomalous changes in the light curve are observed.  
For such a quasar microlensing, hybrid approaches 
are usually employed, where 
particles are located in a smooth gravitational field 
representing a host galaxy. 
It is thus likely that $N$ point-mass approach 
will be useful also when we study such a finite-$N$ effect 
at a certain transit stage 
between a particles system and a smooth one. 
Along this course,  An (2007) investigated a $N$ point lens model, 
which represents a very special configuration that 
every point masses are located on regular grid points.

For a $N$ point-mass lens at a general configuration, 
very few things are known in spite of many efforts. 
Among known ones is 
the maximum number of images lensed by $N$ point masses. 
After direct calculations by Witt (1990) and 
Mao, Petters and Witt (1997), 
a careful study by Rhie 
(2001 for N=4, 2003 for general N) 
revealed that it is possible to obtain 
the maximum number of images as $5(N-1)$. 
This theorem for polynomials has been extended 
to a more general case including rational functions 
by Khavinson and Neumann (2006). 
(See Khavinson and Neumann 2008 for an elegant review 
on a connection between the gravitational lens theory and 
the algebra, especially the fundamental theorem of algebra,  
and its extension to rational functions).  
 
\noindent
{\bf Theorem} (Khavinson and Neumann 2006): \\
Let $r(z)=p(z)/q(z)$, where $p$ and $q$ 
are relatively prime polynomials in $z$, 
and let $n$ be the degree of $r$. 
If $n>1$, then 
the number of zeros for 
$r(z)-z^* \leq 5(n-1)$. 
Here, $z$ and $z^*$ denote a complex number and its complex conjugate, 
respectively. 

Furthermore, Bayer, Dyer and Giang (2006) showed that 
in a configuration of point masses, 
replacing one of the point deflectors 
by a spherically symmetric distributed mass 
only introduces one extra image. 
Hence they found that the maximum number of images 
due to $N$ distributed lensing objects located on a plane
is $6 (N-1)+1$. 

Global properties such as lower bounds on the number of images 
are also discussed 
in Petters, Levine and Wambsganss (2001) and references therein. 

In spite of many efforts on $N$ lensing objects, 
functions for image positions are still unknown 
even for $N$ point-mass lenses in a general configuration 
under the thin lens approximation. 
Hence it is a challenging problem to express 
the image positions as functions of lens and source locations.  
Once such an expression is known,  
one can immediately obtain 
magnifications via computing the Jacobian 
of the lens mapping (Schneider et al. 1992). 

Only for a very few cases such as a single point mass 
and a singular isothermal ellipsoid, 
the lens equation can be solved by hand 
and image positions are known, 
because the lens equation becomes a quadratic or fourth-order one 
(For a singular isothermal ellipsoid, Asada et al. 2003). 
For the binary lens system, the lens equation 
has the degree of five in a complex variable (Witt 1990). 
It has the same degree also in a real variable  
(Asada 2002a, Asada et al. 2004).  
This improvement is not trivial because a complex variable 
brings two degrees of freedom. 
This single-real-variable polynomial has advantages. 
For instance, the number of real roots (with vanishing imaginary
parts) corresponds to that of lensed images. 
The analytic expression of the caustic, where the number of images 
changes, is obtained by the fifth-order polynomial 
(Asada et al. 2002c). 
Galois showed, however,  that the fifth-order and higher 
polynomials cannot be solved algebraically (van der Waerden 1966). 
Hence, no formula for the quintic equation is known. 
For this reason, some numerical implementation 
is required to find out image positions (and magnifications) 
for the binary gravitational lens for a general position 
of the source. 
Only for special cases of the source at a symmetric location 
such as on-axis sources, 
the lens equation can be solved by hand 
and image positions are thus known 
(Schneider and Weiss 1986). 
For a weak field region, some perturbative solutions 
for the binary lens have been found (Bozza 1999, Asada 2002b), 
for instance in order to discuss astrometric lensing,  
which is caused by the image centroid shifts 
(for a single mass, 
Miyamoto and Yoshii 1995, Walker 1995; 
for a binary lens, Safizadeh et al. 1999, Jeong et al. 1999, 
Asada 2002b). 

If the number of point masses $N$ is larger than two, 
the basic equation is much more highly non-linear 
so that the lens equation can be solved 
only by numerical methods. 
As a result, observational properties such 
as magnifications and image separations 
have been investigated so far numerically for 
$N$ point-mass lenses. 
This makes it difficult to investigate 
the dependence of observational quantities 
on lens parameters. 

This paper is the first attempt to seek 
an analytic expression of image positions without 
assuming any special symmetry. 
For this purpose, 
we shall present a method of Taylor-series expansion 
to solve the lens equation for $N$ point-mass lens systems.  
Our method allows a systematic iterative analysis as shown later.  

Under three assumptions of 
weak gravitational fields, thin lenses and small deflection angles, 
gravitational lensing is usually described 
as a mapping from the lens plane onto the source plane 
(Schneider et al. 1992). 
Bourassa and Kantowski (1973, 1975) introduced a complex notation 
to describe gravitational lensing. 
Their notation was exclusively used to 
describe lenses with elliptical or spheroidal symmetry 
(Borgeest 1983, Bray 1984, Schramm 1990). 
For $N$ point lenses, 
Witt (1990) succeeded in recasting the lens equation into 
a single-complex-variable polynomial. 
This is in an elegant form and thus 
has been often used 
in investigations of point-mass lenses. 
An advantage in the single-complex-variable formulation is 
that we can use some mathematical tools applicable to 
complex-analytic functions, especially polynomials 
(Witt 1993, Witt and Petters 1993, Witt and Mao 1995). 
One tool is the fundamental theorem of algebra: 
Every non-constant single-variable polynomial with complex
coefficients has at least one complex root. 
This is also stated as: every non-zero single-variable polynomial, 
with complex coefficients, has exactly as many complex roots 
as its degree, if each root is counted as many times as 
its multiplicity. 
On the other hand, in the original form of the lens equation, 
one can hardly count up the number of images 
because of non-linearly coupled properties.  
This theorem, therefore, raises a problem in gravitational lensing. 
The single-variable polynomial due to $N$ point lenses 
has the degree of $N^2+1$, though the maximum number of 
images is $5(N-1)$. 
This means that unphysical roots are included 
in the polynomial 
(for detailed discussions on the disappearance and appearance 
of images near fold and cusp caustics for general lens systems, 
see also Petters, Levine and Wambsganss (2001) and references therein).  
First, we thus investigate explicitly behaviors of roots 
for the polynomial lens equation from the viewpoint of perturbations. 
We shall identify unphysical roots.  
Secondly, we shall re-examine the lens equation, 
so that the appearance of unphysical roots can be avoided. 
 
This paper is organised as follows.
In Section 2, the complex description 
of gravitational lensing is briefly summarised. 
The lens equation is embedded into a single-complex-variable 
polynomial in Section 3. 
Perturbative roots for the complex polynomial 
are presented for binary and triple systems 
in sections 4 and 5, respectively. 
They are extended to a case of $N$ point lenses 
in section 6. 
In section 7, we re-examine the lens equation 
in a dual-complex-variables formalism 
and its perturbation scheme for 
a binary lens for its simplicity.  
The perturbation scheme is extended to a $N$ point lens system 
in section 8. 
Section 9 is devoted to the conclusion.

\section{Polynomial formalism using complex variables} 
We consider a lens system with N point masses. 
The mass and two-dimensional location of each body 
is denoted as $M_i$ 
and the vector $\mbox{\boldmath $E$}_i$, respectively.  
For the later convenience, let us define the angular size of 
the Einstein ring as 
\begin{equation}
\theta_{E}=
\sqrt{\frac{4GM_{tot} D_{\mbox{LS}}}{c^2 D_{\mbox{L}} D_{\mbox{S}}}} ,  
\end{equation}
where $G$ is the gravitational constant, $c$ is 
the light speed, 
$M_{tot}$ is the total mass $\sum_{i=1}^N M_i$
and $D_{\mbox{L}}$, $D_{\mbox{S}}$ and $D_{\mbox{LS}}$ 
denote distances 
between the observer and the lens, 
between the observer and the source, and 
between the lens and the source, respectively. 
In the unit normalised by the angular size of the Einstein ring, 
the lens equation becomes 
\begin{equation}
\mbox{\boldmath $\beta$}=\mbox{\boldmath $\theta$}
-\sum_i^N  
\nu_i  
\frac{\mbox{\boldmath $\theta$}-\mbox{\boldmath $e$}_i}
{|\mbox{\boldmath $\theta$}-\mbox{\boldmath $e$}_i|^2} , 
\label{lenseq}
\end{equation}
where 
$\mbox{\boldmath $\beta$} = (\beta_x, \beta_y)$ 
and 
$\mbox{\boldmath $\theta$} = (\theta_x, \theta_y)$ 
denote the vectors for the position of the source and image, 
respectively and we defined the mass ratio and the angular 
separation vector as 
$\nu_i = M_i/M_{tot}$ 
and 
$\mbox{\boldmath $e$}_i = \mbox{\boldmath $E$}_i/\theta_{E} 
= (e_{ix}, e_{iy})$. 

\begin{figure}
\includegraphics[width=84mm]{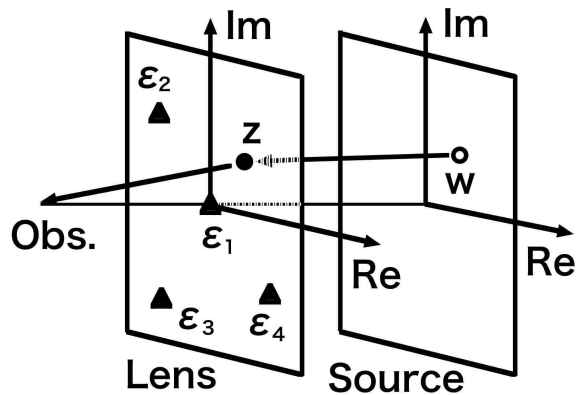}
\caption{
Notation: 
The source and image positions on complex planes are denoted 
by $w$ (the circle) and $z$ (the filled disk), 
respectively. 
Locations of N point masses are denoted by $\epsilon_i$  
(filled triangles) for $i=1, \cdots, N$. 
Here, we assume the thin lens approximation. 
}
\label{notation}
\end{figure}

In a formalism based on complex variables, 
two-dimensional vectors for the source, lens and image positions   
are denoted as $w=\beta_x+i \beta_y$, 
$z = \theta_x + i \theta_y$,  
and 
$\epsilon_i=e_{ix} + i e_{iy}$, 
respectively (See also Fig. $\ref{notation}$).   
By employing this formalism, the lens equation is rewritten as 
\begin{equation}
w =z - \sum_i^N \frac{\nu_i}{z^* - \epsilon_i^*} , 
\label{lenseq-z}
\end{equation}
where the asterisk $*$ means the complex conjugate.  
The lens equation is non-analytic because 
it contains both $z$ and $z^*$. 


\section{Embedding the lens equation into an analytic polynomial} 
The complex conjugate of Eq. ($\ref{lenseq-z}$) 
is expressed as 
\begin{equation}
w^* =z^* - \sum_i^N \frac{\nu_i}{z - \epsilon_i} .  
\label{lenseq-zcc}
\end{equation}
This expression can be substituted into $z^*$ in Eq. ($\ref{lenseq-z}$) 
to eliminate the complex variable $z^*$. 
As a result, we obtain a $(N^2+1)$-th order analytic polynomial equation as 
(Witt 1990)
\begin{eqnarray}
&&(z-w) \prod_{\ell=1}^N 
\left(
(w^*-\epsilon_{\ell}^*) \prod_{k=1}^N (z-\epsilon_k)
+ \sum_{k=1}^N \nu_k \prod_{j \neq k}^N (z-\epsilon_j) 
\right) 
\nonumber\\
&&= \sum_{i=1}^N \nu_i \prod_{\ell=1}^N(z-\epsilon_{\ell}) 
\nonumber\\
&&\times 
\prod_{m \neq i}^N 
\left(
(w^* - \epsilon_m^*) \prod_{k=1}^N (z-\epsilon_k) 
+ \sum_{k=1}^N \nu_k 
\prod_{j \neq k}^N (z-\epsilon_j) 
\right) . 
\nonumber\\
&&
\label{N-lenseq}
\end{eqnarray} 
Equation (A3) in Witt (1990) takes a rather complicated form 
because of inclusion of nonzero shear $\gamma$ 
due to surrounding matter. 
Bayer et al. (2006) uses a complex formalism in order to 
discuss the maximum number of images 
in a configuration of point masses, by replacing one 
of point deflectors by a spherically symmetric distributed mass. 
Their lens equation (3) 
agrees with Eq. ($\ref{N-lenseq}$). 
In order to show this agreement, 
one may use $(-1)^{N+1}=(-1)^{N-1}$. 
It is worthwhile to mention that Eq. $(\ref{N-lenseq})$ 
contains not only all the solutions for 
the lens equation $(\ref{lenseq})$ but also 
unphysical false roots which do not satisfy 
Eq. $(\ref{lenseq})$, 
in price of the manipulation for obtaining 
an analytic polynomial equation, 
as already pointed out 
by Rhie (2001, 2003) and Bayer et al. (2006). 
Such an inclusion of unphysical solutions 
can be easily understood by remembering 
that we get unphysical roots as well as true ones 
if one takes a square of an equation including the square root.  
In fact, an analogous thing happens in another example 
of gravitational lenses such as an isothermal ellipsoidal 
lens as a simple model of galaxies (Asada et al. 2003). 

In general, the mass ratio $\nu_i$ satisfies 
$0<\nu_i<1$, so that it can be taken as 
an expansion parameter. 
Without loss of generality, we can assume that 
the first lens object is the most massive, namely 
$m_1 \geq m_i$ for $i= 2, 3, \cdots, N$. 
Thus, formal solutions are expressed in Taylor series as
\begin{equation}
z=\sum_{p_2=0}^{\infty}\sum_{p_3=0}^{\infty} \cdots
\sum_{p_N=0}^{\infty}
\nu_2^{p_2}\nu_3^{p_3}\cdots \nu_N^{p_N} 
z_{(p_2)(p_3)\cdots (p_N)} , 
\label{formalsolution}
\end{equation}
where the coefficients $z_{(p_2)(p_3)\cdots (p_N)}$ 
are independent of $\nu_i$. 

Up to this point, the origin of the lens plane 
is arbitrary. 
In the following, the origin of the lens plane is 
chosen as the location of the mass $m_1$, 
such that one can put $\epsilon_1 =0$. 
This enables us to simplify some expressions and 
to easily understand their physical meanings, 
mostly because gravity is dominated by $m_1$ 
in most regions except for the vicinity of $m_i$ $(i\neq 1)$. 
Namely, it is natural to treat our problem as 
perturbations around a single lens by $m_1$ 
(located at the origin of the coordinates). 

In numerical simulations or practical data analysis, 
however, one may use the coordinates in which the origin 
is not the location of $m_1$. 
If one wishes to consider such a case of $\epsilon_1 \neq 0$, 
one could make a translation by $\epsilon_1$ as 
$z \to z+\epsilon_1$, $w \to w+\epsilon_1$ 
and $\epsilon_i \to \epsilon_i +\epsilon_1$ 
in our perturbative solutions that are given below.

\section{Perturbative Solutions For A Polynomial Formalism 1: Binary Lens}
In this section, we investigate binary lenses explicitly 
up to the third order. 
This simple example may help us to understand the structure 
of the perturbative solutions. 
For an arbitrary $N$ case, expressions of iterative solutions 
are quite formal (See below). 

For simplicity, we denote our expansion parameter as 
$m \equiv \nu_2$. 
This means $\nu_1 = 1-m$. 
We also denote $\epsilon_2$ simply by $\epsilon$. 

In powers of $m$, the polynomial equation is 
rewritten as 
\begin{equation}
\sum_{k=0}^2 m^k f_k(z) = 0 ,
\end{equation}
where we defined 
\begin{eqnarray}
f_0(z)&=&(z-\epsilon)^2 [(w^*-\epsilon^*)z+1] 
(w^* z^2-w w^* z - w) ,
\nonumber\\
f_1(z)&=&(z-w) 
\nonumber\\
&&
\times 
\Bigl(
\epsilon (z-w) [(2w^*-\epsilon^*)z+2] 
-\epsilon^* z^2 (z-\epsilon) - \epsilon z 
\Bigr) , 
\nonumber\\
f_2(z)&=&\epsilon^2 (z-w) . 
\label{f}
\end{eqnarray}

We seek a solution in expansion series as 
\begin{equation}
z=\sum_{p=0}^{\infty} m^p z_{(p)} .  
\end{equation}

\subsection{0th order}
At $O(m^0)$, the lens equation becomes 
the fifth-order polynomial equation as $f_0 = 0$. 
Zeroth order solutions are obtained by solving this. 
All the solutions are 
$\epsilon$ (doublet), $\alpha_3$   
and $\alpha_{\pm}$,  
where we defined 
\begin{eqnarray}
\alpha_3 &=& \frac{1}{\epsilon^*-w^*} ,  
\nonumber\\
\alpha_{\pm} &=& \frac{w}{2} 
\left(1 \pm \sqrt{1+\frac{4}{w w^*}}\right) . 
\label{2-sol-0th}
\end{eqnarray}
One of the roots, $\alpha_3$,  is unphysical, 
because it does not satisfy Eq. $(\ref{lenseq})$ at $O(m^0)$. 
By using all the 0th order roots including unphysical ones, 
$f_0$ is factorised as 
\begin{equation}
f_0(z) = w^* (w^*-\epsilon^*) (z-\epsilon)^2 (z-\alpha_3) 
(z-\alpha_+) (z-\alpha_-) . 
\end{equation}

\subsection{1st order}
Next, we seek $1$st-order roots. 
We put $z=\alpha_{\pm} + m z_{(1)} +O(m^2)$. 
At the linear order in $m$, Eq. ($\ref{N-lenseq}$) becomes 
\begin{equation}
z_{(1)} f_0^{'}(\alpha_{\pm}) + f_1(\alpha_{\pm}) = 0 , 
\end{equation}
where the prime denotes the derivative with respect to $z$. 
Thereby we obtain a $1$st-order root as 
\begin{equation}
z_{(1)} = -\frac{f_1(\alpha_{\pm})}{f_0^{'}(\alpha_{\pm})} . 
\end{equation}
The similar manner cannot be applied to a case of $\epsilon$, 
because it is a doublet root with 
$f_0(\epsilon) = f_0^{'}(\epsilon) = 0$, while 
$f_0^{''}(\epsilon) \neq 0$. 
At $O(m^2)$, Eq. ($\ref{N-lenseq}$) can be factorised as 
\begin{eqnarray}
&&
\left(z_{(1)} [(w^*-\epsilon^*)\epsilon +1] + \epsilon \right)
\nonumber\\
&&
\times 
\left(z_{(1)} [(\epsilon-w)(w^* \epsilon + 1)-\epsilon] 
+ \epsilon (\epsilon-w) \right)
= 0 . 
\label{2-lenseq-2nd}
\end{eqnarray} 
Hence, we obtain two roots as 
\begin{eqnarray}
z_{(1)}&=&\frac{\epsilon}{(\epsilon^*-w^*)\epsilon - 1} , 
\label{z1-1}
\\
z_{(1)}&=& - \frac{\epsilon (\epsilon - w)}
{(\epsilon - w) (w^* \epsilon +1) - \epsilon} .  
\label{z1-2}
\end{eqnarray}
Here, the latter root expressed by Eq. ($\ref{z1-2}$) 
is unphysical and thus abandoned, because 
it doesn't satisfy the original lens equation ($\ref{lenseq}$).  
On the other hand,  
the former root by Eq. ($\ref{z1-1}$) satisfies the equation 
and thus expresses a physically correct image.

\subsection{2nd Order}
First, we consider perturbations around zeroth-order solutions 
of $\alpha_{\pm}$. 
At $O(m^2)$, Eq. ($\ref{N-lenseq}$) is linear in $z_{(2)}$ 
and thus easily solved for $z_{(2)}$ as 
\begin{equation}
z_{(2)} = 
- \frac
{z_{(1)}^2 f_0^{''}(\alpha_{\pm}) +2 z_{(1)} f_1^{'}(\alpha_{\pm}) 
+2 f_2(\alpha_{\pm})} 
{2 f_0^{'}(\alpha_{\pm})} . 
\label{polynomial-binary-z2-1}
\end{equation}

Next, we investigate a multiple root $\epsilon$. 
At $O(m^3)$, Eq. ($\ref{N-lenseq}$) becomes linear in $z_{(2)}$. 
It is solved as 
\begin{equation}
z_{(2)} = 
- \frac
{z_{(1)}^3 f_0^{'''}(\epsilon) + 3 z_{(1)}^2 f_1^{''}(\epsilon) 
+6 z_{(1)} f_2^{'}(\epsilon)} 
{6 [z_{(1)} f_0^{''}(\epsilon) + f_1^{'}(\epsilon)]} . 
\label{polynomial-binary-z2-2}
\end{equation}

\subsection{3rd Order}
Around zeroth-order solutions of $\alpha_{\pm}$, 
Eq. ($\ref{N-lenseq}$) at $O(m^3)$ is linear in $z_{(2)}$ 
and thus solved as 
\begin{eqnarray}
z_{(3)} 
&=& 
- \frac{1}{f_0^{'}(\alpha_{\pm})}
[z_{(1)} z_{(2)} f_0^{''}(\alpha_{\pm}) 
+ \frac16 z_{(1)}^3 f_0^{'''}(\alpha_{\pm}) 
\nonumber\\
&&
+ z_{(2)} f_1^{'}(\alpha_{\pm}) 
+ \frac12 z_{(1)}^2 f_1^{''}(\alpha_{\pm}) 
+ z_{(1)} f_2^{'}(\alpha_{\pm}) ] .
\end{eqnarray}

Also around the multiple root $\epsilon$, 
Eq. ($\ref{N-lenseq}$) at $O(m^3)$ becomes linear 
in $z_{(2)}$. 
It is solved as 
\begin{eqnarray}
z_{(3)} 
&=&  
- \frac{1}{z_{(1)} f_0^{''}(\epsilon) + f_1^{'}(\epsilon)} 
\nonumber\\
&&
\times
[\frac12 z_{(2)}^2 f_0^{''}(\epsilon) 
+\frac12 z_{(1)}^2 z_{(2)} f_0^{'''}(\epsilon) 
+\frac{1}{24} z_{(1)}^4 f_0^{''''}(\epsilon)
\nonumber\\
&&
+ z_{(1)} z_{(2)} f_1^{''}(\epsilon)
+ \frac16 z_{(1)}^3 f_1^{'''}(\epsilon) 
\nonumber\\
&&
+2 z_{(2)} f_2^{'}(\epsilon)] ,  
\end{eqnarray}
where we used $f_2^{''}(z)=0$. 

Table $\ref{table-polynomial}$ shows a numerical example 
of perturbative roots and their convergence. 


\begin{table*}
\begin{minipage}{180mm}
\caption{
Example of perturbative roots 
in the single-complex-polynomial: 
We assume $\nu=0.1$, $e=1$ and two cases of 
$w=2$ (on-axis) and $w=1+i$ (off-axis). 
In this example, the number of images is three. 
''Polynomial'' in the table means all the five roots 
for the single-complex-polynomial. 
They are obtained by numerically solving the polynomial. 
Image positions are determined also by numerically solving 
the lens equation. They are listed in the column 
''Lens Eq.''. 
 In the table, ''None'' means that 
it does not exist. 
These tables show that, as we go to higher orders, 
the perturbative roots become closer to the correct ones 
for the single-complex-polynomial, 
including the two unphysical roots. 
}
\begin{center}
    \begin{tabular}{llllll}
\hline
Case 1 & On-axis & $\nu=0.1$ &  $e=1$ & $w=2$ 
\\
\hline
Root & 1 & 2 & 3 & 4 & 5 
\\
\hline
1st. & 2.43921 & -0.389214 & 0.95 & -0.925 & 0.975 
\\
2nd. & 2.43855 & -0.388551 & 0.95 & -0.924063 & 0.974063
\\
3rd. & 2.43858 & -0.388519 & 0.949938 & -0.924016 & 0.974016
\\
\hline
Polynomial & 2.43858 & -0.388517 & 0.949937 & -0.924013 & 0.974013 
\\
\hline
Lens Eq. & 2.43858 & -0.388517 & 0.949937 & None & None 
\\
\hline
    \end{tabular}
  \end{center}

\begin{center}
    \begin{tabular}{llllll}
\hline
Case 2 & Off-axis & $\nu=0.1$ &  $e=1$ & $w=1+i$ 
\\
\hline
Root & 1 & 2 & 3 & 4 & 5 
\\
\hline
1st. & 1.33716+1.40546 i & -0.337158-0.355459 i & 0.95-0.05 i 
& 0.025-0.925 i & 0.975-0.025 i 
\\
2nd. & 1.33632+1.40363 i & -0.336316-0.354881 i & 0.95-0.05 i 
& 0.02625-0.9225 i & 0.97375-0.02625 i
\\
3rd. & 1.33634+1.40371 i & -0.336275-0.354839 i & 0.95-0.05025 i 
& 0.0262813-0.922281 i & 0.973656-0.0263438 i 
\\
\hline
Polynomial & 1.33633+1.40371 i & -0.336272-0.354835 i  
& 0.950015-0.0502659 i & 0.0262762-0.922254 i & 0.973646-0.0263517 i
\\
\hline
Lens Eq. & 1.33633+1.40371 i & -0.336272-0.354835 i 
& 0.950015-0.0502659 i & None & None 
\\
\hline
    \end{tabular}
  \end{center}
\label{table-polynomial}
\end{minipage}
\end{table*}

\section{Perturbative Solutions For A Polynomial Formalism 2: Triplet Lens}
In a binary case, we have only the single parameter $m$ 
for the perturbations. 
For $N$ point masses, we have to take account of couplings 
among several expansion parameters. 
In addition, the degree of the polynomial becomes $N^2+1$, 
so that we cannot write down the whole equation. 
In order to get hints for $N$ point-mass lenses, 
in this section, we investigate triple-mass lenses explicitly 
up to the second order in $\nu_2$ and $\nu_3$. 

The polynomial equation is rewritten as  
\begin{equation}
\sum_{p_2=0}^3\sum_{p_3=0}^3 (\nu_2)^{p_2} (\nu_3)^{p_3} g_{(p_2)(p_3)}(z) = 0 ,
\end{equation}
where we defined 
\begin{eqnarray}
g_{(0)(0)}(z)&=&(z-\epsilon_2)^3 (z-\epsilon_3)^3 
[(w^*-\epsilon_2^*)z+1] 
\nonumber\\
&&
\times 
[(w^*-\epsilon_3^*)z+1]
(w^* z^2-w w^* z - w) . 
\label{f}
\end{eqnarray}

We seek a solution in expansion series as 
\begin{equation}
z=\sum_{p_2=0}^{\infty} \sum_{p_3=0}^{\infty} 
(\nu_2)^{p_2} (\nu_3)^{p_3} z_{(p_2)(p_3)} .  
\end{equation}

\subsection{0th order}
Zeroth order solutions are obtained by solving 
the tenth-order polynomial equation as $g_{(0)(0)} = 0$. 
The roots are 
$\epsilon_2$ (doublet), $\epsilon_3$ (doublet), $\alpha_3$,   
$\alpha_4$ and $\alpha_{\pm}$,  
where we defined 
\begin{eqnarray}
\alpha_3 &=& \frac{1}{\epsilon_2^*-w^*}  , 
\nonumber\\
\alpha_4 &=& \frac{1}{\epsilon_3^*-w^*}  . 
\label{2-sol-0th}
\end{eqnarray} 
For the same reason in the binary lens, 
$\alpha_3$ and $\alpha_4$ are unphysical, 
in the sense that it does not satisfy 
the lens equation $(\ref{lenseq})$. 
By using all the 0th order roots, $g_{(0)(0)}$ is factorised as 
\begin{eqnarray}
g_{(0)(0)}(z) &=& w^* (w^*-\epsilon_2^*) (w^*-\epsilon_3^*) 
(z-\epsilon_2)^3 (z-\epsilon_3)^3 
\nonumber\\
&&
\times
(z-\alpha_3) (z-\alpha_4)
(z-\alpha_+) (z-\alpha_-) . 
\end{eqnarray}

\subsection{1st order}
Here, we seek $1$st-order roots. 
The image position is expanded as 
$z=\alpha_{\pm} + \nu_2 z_{(1)(0)} + \nu_3 z_{(0)(1)} 
+ O(\nu_2^2, \nu_3^2, \nu_2 \nu_3)$. 
By making a replacement in notations as $2 \leftrightarrow 3$, 
one can construct $z_{(0)(1)}$ from $z_{(1)(0)}$. 
Hence, we focus on $z_{(1)(0)}$ below. 
At the linear order in $\nu_2$, 
Eq. ($\ref{N-lenseq}$) becomes 
\begin{equation}
z_{(1)(0)} g_{(0)(0)}^{'}(\alpha_{\pm}) 
+ g_{(1)(0)}(\alpha_{\pm}) = 0 .  
\end{equation} 
Thereby we obtain a $1$st-order root as 
\begin{equation}
z_{(1)(0)} = -\frac{g_{(1)(0)}(\alpha_{\pm})}
{g_{(0)(0)}^{'}(\alpha_{\pm})} . 
\end{equation}

For the triple mass lens system, the root $\epsilon_2$ 
becomes triplet with 
$g_{(00)}(\epsilon_2) = g_{(00)}^{'}(\epsilon_2) = 
g_{(00)}^{''}(\epsilon_2) = 0$, while 
$g_{(00)}^{'''}(\epsilon) \neq 0$. 
After rather lengthy but straightforward calculations,
Eq. ($\ref{N-lenseq}$) at $O(\nu_2^2)$ can be factorised as 
\begin{eqnarray}
&&
\left(z_{(1)(0)} [(w^*-\epsilon_2^*)\epsilon_2 +1] + \epsilon_2 \right)
\nonumber\\
&&
\times
\left(z_{(1)(0)} [(w^*-\epsilon_3^*)\epsilon_2 +1] + \epsilon_2 \right)
\nonumber\\
&&
\times 
\left(z_{(1)(0)} [(\epsilon_2-w)(w^* \epsilon_2 + 1)-\epsilon_2] 
+ \epsilon_2 (\epsilon_2-w) \right) 
\nonumber\\
&&=0 . 
\label{2-lenseq-2nd}
\end{eqnarray} 
Hence, we obtain three roots as 
\begin{eqnarray}
z_{(1)(0)}&=&\frac{\epsilon_2}{(\epsilon_2^*-w^*)\epsilon_2 - 1} , 
\label{z10-1}
\\
z_{(1)(0)}&=&\frac{\epsilon_2}{(\epsilon_3^*-w^*)\epsilon_2 - 1} , 
\label{z10-2}
\\z_{(1)(0)}&=&- \frac{\epsilon_2 (\epsilon_2 - w)}
{(\epsilon_2 - w) (w^* \epsilon_2 +1) - \epsilon_2} . 
\label{z10-3}
\end{eqnarray}
At the linear order in $\nu_2$, true solutions for the triple 
lens system has to agree with that for the binary system, 
when one takes a limit as $\nu_3 \to 0$. 
Therefore, out of the above three roots, ones expressed by 
Eqs. ($\ref{z10-2}$) and ($\ref{z10-3}$) must be abandoned.

\section{Perturbative Solutions For A Polynomial Formalism 3: 
N Point-Mass Lens}
In the previous section, we have learned couplings 
between the second and third masses. 
Now we are in a position to investigate a lens system 
consisting of N point masses. 

The polynomial lens equation ($\ref{N-lenseq}$) is expanded as  
\begin{eqnarray}
&&\sum_{p_2=0}^N \sum_{p_3=0}^N \cdots \sum_{p_N=0}^N
(\nu_2)^{p_2} (\nu_3)^{p_3} \cdots (\nu_N)^{p_N} 
\nonumber\\
&&
\times 
g_{(p_2)(p_3) \cdots (p_N)}(z) 
= 0 . 
\end{eqnarray}
For this equation, 
we seek a solution in expansion series as 
\begin{eqnarray}
z&=&\sum_{p_2=0}^{\infty} \sum_{p_3=0}^{\infty} \cdots
\sum_{p_N=0}^{\infty} 
(\nu_2)^{p_2} (\nu_3)^{p_3} \cdots (\nu_N)^{p_N} 
\nonumber\\
&&
\times 
z_{(p_2)(p_3) \cdots (p_N)} .
\end{eqnarray}

\subsection{0th order}
Zeroth order solutions are obtained by solving 
the $(N^2+1)$th-order polynomial equation as 
$g_{(0) \cdots (0)} = 0$. 
The roots are 
$\alpha_i \equiv -1/w_i^*$, $\alpha_{\pm}$, and  
$\epsilon_i$ (with multiplicity = $N$) for $i=2, \cdots N$,
where for later convenience we denoted 
\begin{equation}
w_i = w - \epsilon_i . 
\end{equation} 
Like in the binary lens, 
$\alpha_i$ is unphysical, 
in the sense that it does not satisfy 
the lens equation $(\ref{lenseq})$. 
By using all the 0th order roots, 
$g_{(0) \cdots (0)}$ is factorised as 
\begin{eqnarray}
g_{(0)\cdots  (0)}(z) 
&=& 
(z-\alpha_+) (z-\alpha_-) 
w^* 
\prod_{j=2}^N (w_j^*) 
\nonumber\\
&&
\times
\prod_{k=2}^N (z-\epsilon_k)^N 
\prod_{\ell=2}^N (z+\frac{1}{w_{\ell}^*}) , 
\end{eqnarray}
where this degree is $N^2+1$ in agreement with
that of the polynomial equation.

\subsection{1st order}
Next, we seek $1$st-order roots. 
In the similar manner in the double or triple mass case, 
we can obtain a $1$st-order root as 
\begin{equation}
z_{(0) \cdots (1_k) \cdots (0)} 
= -\frac{g_{(0) \cdots (1_k) \cdots (0)}(\alpha_{\pm})}
{g_{(0) \cdots (0)}^{'}(\alpha_{\pm})} , 
\end{equation}
where $1_k$ denotes that the $k$-th index is the unity, 
namely $p_k = 1$.  

For $N$ point mass lens systems, a root $\epsilon_k$ 
is multiplet with multiplicity = $N$. 
Without loss of generality, we choose $\epsilon_2$ 
as a root in the following discussion. 
Calculations done above for a double or triple mass system 
suggest that Eq. ($\ref{N-lenseq}$) at $O(\nu_2^2)$ 
can be factorised as 
\begin{eqnarray}
&&
\prod_{k=2}^N
\left(z_{(1)(0) \cdots (0)} [(w_k^*)\epsilon_2 +1] + \epsilon_2 \right)
\nonumber\\
&&
\times 
\left(z_{(1)(0) \cdots (0)} [w_2 (w^* \epsilon_2 + 1)+\epsilon_2] 
+ \epsilon_2 w_2 \right)
= 0 . 
\label{N-lenseq-2nd}
\end{eqnarray} 
By using this factorisation, we obtain $N$ roots. 
At the linear order in $\nu_2$, however, 
true solutions for the present lens system has to agree with 
that for the binary system, 
because one can take the limit as $\nu_p \to 0$ for $p \geq 3$. 
Therefore, only the $-\epsilon_2/(w_2^* \epsilon_2 +1)$ 
out of the above $N$ roots is correct for 
the original lens equation.  
The same argument is true of any $\epsilon_i$.

\section{Perturbative Solutions For $zz^*$-dual formalism 1: Binary Lens}
As shown above, an analytic polynomial formalism is apparently simple. 
When we solve perturbatively the polynomial equation, however, 
we find unphysical roots 
which satisfy the polynomial but does not the original lens equation. 
In the polynomial formalism, therefore, we are required to 
check every roots and then to pick up only the physical roots 
satisfying the original lens equation 
with discarding unphysical ones. 
It is even worse that the order of the polynomial grows rapidly 
as $N^2+1$, as the number of the lens objects increases. 
This means that the perturbative structure of the formalism 
becomes much more complicated as $N$ increases. 
In this section, we thus investigate 
another formalism, which allows 
a more straightforward calculation 
especially without needing extra procedures such as deleting 
physically incorrect roots. 

First, we focus on a binary case for its simplicity. 
The lens equation is rewritten as 
\begin{equation}
C(z, z^*) = m D(z^*) , 
\label{C=mD}
\end{equation}
where we defined 
\begin{eqnarray}
C(z, z^*)&=&w-z+\frac{1}{z^*} , 
\label{C}
\\
D(z^*)&=&\frac{1}{z^*}-\frac{1}{z^*-\epsilon^*} . 
\label{D}
\end{eqnarray}

One of advantages in this $zz^*$-formulation is that 
the master equation $(\ref{C=mD})$ is linear in $m$. 
Therefore, counting orders in $m$ 
can be drastically simplified 
when we perform iterative calculations. 
On the other hand, an analytic polynomial is 
second order in $m$. 
In fact, practical perturbative computations 
in the polynomial formalism are quite complicated, 
in the sense that 
several different terms ($f_0$, $f_1$ and $f_2$ for a binary case) 
may make the same order-of-magnitude contributions 
at each iteration step. 

We seek a solution in expansion series as 
\begin{equation}
z=\sum_{k=0}^{\infty} m^k z_{(k)} .  
\label{z-exp}
\end{equation}
The complex conjugate of this becomes  
\begin{equation}
z^*=\sum_{k=0}^{\infty} m^k z_{(k)}^* .  
\label{z*-exp}
\end{equation}
According to these power-series expansions of $z$ and $z^*$, 
both sides of the lens equation are expanded as 
\begin{equation}
C(z, z^*) = \sum_{k=0}^{\infty} m^k C_{(k)} ,   
\label{G-exp}
\end{equation}
\begin{equation}
D(z^*) = \sum_{k=0}^{\infty} m^k D_{(k)} , 
\label{H-exp}
\end{equation}
where $C_{(k)}$ and $D_{(k)}$ are independent of $m$. 
At $O(m^k)$, Eq. ($\ref{C=mD}$) becomes 
\begin{equation}
C_{(k)} = D_{(k-1)} , 
\end{equation}
which shows clearly a much simpler structure than 
a polynomial case such as 
Eqs. ($\ref{polynomial-binary-z2-1}$) 
and ($\ref{polynomial-binary-z2-2}$).   
Equation ($\ref{D}$) indicates that 
$D(z^*)$ has a pole at $z^* = \epsilon^*$. 
Therefore, we shall discuss two cases of 
$z_{(0)} \neq \epsilon$ and $z_{(0)} = \epsilon$, separately.

\subsection{0th order ($z_{(0)} \neq \epsilon$)}
Zeroth order solutions are obtained by solving 
the equation as 
\begin{equation}
C(z_{(0)}, z^*_{(0)}) = 0.
\label{binary-C-0}
\end{equation} 
The solution for this is the well-known roots 
for a single mass lens. 
In order to help readers to understand 
the $zz^*$-dual formulation, we shall derive the roots 
by keeping both $z$ and $z^*$.  
In conventional treatments, 
a single lens case is reduced to one-dimensional one 
by choosing the source direction along the $x$-axis 
in vector formulations or the real axis in complex ones. 
Eq. $(\ref{binary-C-0})$ is rewritten as 
\begin{equation}
z_{(0)} z_{(0)}^* - 1 = w z_{(0)}^*. 
\label{2-G1}
\end{equation} 
The L. H. S. is purely real so that 
the R. H. S. must be real. 
Unless $w = 0$, therefore, one can put $z_{(0)}=A w$ 
by introducing a certain real number $A$. 
By substituting $z_{(0)}=A w$ into Eq. ($\ref{2-G1}$), 
one obtains a quadratic equation for $A$ as 
\begin{equation}
w w^* A^2 - w w^* A - 1 = 0 . 
\label{2-G2}
\end{equation} 
This is solved as 
\begin{equation}
A = \frac12 
\left(1 \pm \sqrt{1+\frac{4}{w w^*}}\right) ,  
\label{2-G3}
\end{equation}
which gives the 0th-order solution. 

In the special case of $w=0$, Eq. ($\ref{2-G1}$) becomes 
$|z_{(0)}| = 1$, which is the Einstein ring. 
In the following, we assume $w \neq 0$. 

\subsection{1st order ($z_{(0)} \neq \epsilon$)}
In units of $z_{(0)}$, 
the expansion series of $z$ is normalised as 
\begin{equation}
z=z_{(0)} \sum_{k=0}^{\infty} m^k \sigma_{(k)} , 
\label{z-exp-2}
\end{equation}
where we defined $\sigma_{(k)} = z_{(k)} / z_{(0)}$. 

First, we investigate a case of $z_{(0)} \neq \epsilon$.  
At the linear order in $m$, Eq. ($\ref{C=mD}$) becomes 
\begin{equation}
z_{(1)} + \frac{z_{(1)}^*}{(z_{(0)}^*)^2} 
= - \frac{1}{z_{(0)}^*} + \frac{1}{z_{(0)}^* - \epsilon^*} .
\label{binary-zz*1st}
\end{equation}

In order to solve Eq. ($\ref{binary-zz*1st}$), 
we consider an equation linear in both $z$ and $z^*$ as 
\begin{equation}
z + a z^* = b ,
\label{zz*ab}
\end{equation}
for two complex constants $a, b \in C$.

Unless $|a|=1$, 
the general root for this equation is 
\begin{equation}
z = \frac{b-ab^*}{1-aa^*} . 
\label{zz*root}
\end{equation}
This can be verified by a direct substitution of 
Eq. $(\ref{zz*root})$ into Eq. $(\ref{zz*ab})$. 
If $|a|=1$, Eq. $(\ref{zz*ab})$ is underdetermined,  
in the sense that it could not provide the unique root without any 
additional constraint condition on $z$ and $z^*$. 

By using directly Eq. ($\ref{zz*root}$), Eq. ($\ref{binary-zz*1st}$) 
is solved as 
\begin{eqnarray}
z_{(1)} = 
\frac{1}{z_{(0)}^2 (z_{(0)}^*)^2 -1} 
\left(
\frac{\epsilon^* z_{(0)}^2 z_{(0)}^*}{z_{(0)}^* - \epsilon^*} 
- \frac{\epsilon z_{(0)}}{z_{(0)} - \epsilon}
\right) . 
\end{eqnarray}

\subsection{2nd order ($z_{(0)} \neq \epsilon$)}
At $O(m^2)$, Eq. ($\ref{C=mD}$) is 
\begin{equation}
z_{(2)} + a_2 z_{(2)}^* 
= b_2 , 
\label{binary-zz*2nd}
\end{equation}
where we defined 
\begin{eqnarray}
a_2&=&\frac{1}{(z_{(0)}^*)^2} , 
\label{a2}
\\
b_2&=& -D_{(1)}  
+ \frac{(\sigma_{(1)}^*)^2}{z_{(0)}^*} . 
\label{b2}
\end{eqnarray}
Here, $D_{(1)}$ is written as 
\begin{equation}
D_{(1)} = 
-\frac{\sigma_{(1)}^*}{z_{(0)}^*} 
+\frac{\sigma_{(1)}^*}{z_{(0)}^*-\epsilon^*} . 
\end{equation}

By using the relation ($\ref{zz*root}$), 
Eq. ($\ref{binary-zz*2nd}$) is solved as 
\begin{equation}
z_{(2)} = \frac{b_2-a_2 b_2^*}{1-a_2 a_2^*} . 
\end{equation}

\subsection{3rd order and $n$th order ($z_{(0)} \neq \epsilon$)}
Computations at $O(m^3)$ are similar to those at $O(m^2)$ 
as shown below.  
At $O(m^3)$, 
Eq. ($\ref{C=mD}$) takes a form as 
\begin{equation}
z_{(3)} + a_3 z_{(3)}^* 
= b_3 , 
\label{binary-zz*3rd}
\end{equation}
where we defined 
\begin{eqnarray}
a_3&=&\frac{1}{(z_{(0)}^*)^2} , 
\label{a3}
\\
b_3&=& -D_{(2)}  
+ \frac{2 \sigma_{(1)}^* \sigma_{(2)}^* - (\sigma_{(1)}^*)^3}
{z_{(0)}^*} . 
\label{b3}
\end{eqnarray}
Here, $D_{(2)}$ is written as 
\begin{equation}
D_{(2)} = 
-\frac{\sigma_{(2)}^* - (\sigma_{(1)}^*)^2}{z_{(0)}^*} 
+\frac{z_{(2)}^*}{(z_{(0)}^*-\epsilon^*)^2}
-\frac{(z_{(1)}^*)^2}{(z_{(0)}^*-\epsilon^*)^3} . 
\end{equation}

Using the relation ($\ref{zz*root}$) 
for Eq. ($\ref{binary-zz*3rd}$),  
we obtain  
\begin{equation}
z_{(3)} = \frac{b_3-a_3 b_3^*}{1-a_3 a_3^*} . 
\end{equation}

In the similar manner, one can obtain 
iteratively $n$th-order roots $z_{(n)}$, 
which obeys an equation in the form of 
Eq. $(\ref{zz*ab})$, 
and thus can use Eq. $(\ref{zz*root})$ to obtain $z_{(n)}$.

\subsection{0th and 1st order ($z_{(0)} = \epsilon$)} 
Next, we investigate the vicinity of $z = \epsilon$, 
which is a pole of $D$. 
The other pole of $D$ is $z=0$, which makes also 
$C(z, z*)$ divergent. Therefore, $z=0$ and 
its neighbourhood are abandoned. 
Let us focus on a root around $z = \epsilon$. 

We assume $z=\epsilon + m z_{(1)} +O(m^2)$. 
Then, the relevant terms in expansion series 
of $C$ and $D$ become  
\begin{eqnarray}
C_{(0)}&=&w-\epsilon+\frac{1}{\epsilon^*} , 
\\
D_{(-1)}&=&- \frac{1}{z_{(1)}^*} , 
\end{eqnarray}
where the index $-1$ means that 
the inverse of $m$ appears because of the pole at $\epsilon$. 
Therefore, the lens equation at $O(m^0)$ becomes 
linear in $z_{(1)}^*$ without including $z_{(1)}$.  
Immediately, it determines $z_{(1)}^*$.  
Its complex conjugate becomes 
\begin{equation}
z_{(1)} = - \frac{\epsilon}{(w^* - \epsilon^*) \epsilon + 1} . 
\end{equation}
This shows a clear difference between 
$z_{(0)} = \epsilon$ and $z_{(0)} \neq \epsilon$ cases. 
Equation ($\ref{binary-zz*1st}$) for the latter case contains 
both $z_{(1)}$ and $z_{(1)}^*$,   
so that we must use a relation such as Eq. $(\ref{zz*root})$.

\subsection{2nd, 3rd and $n$th order ($z_{(0)} = \epsilon$)} 
Next, we consider the lens equation at $O(m^1)$, 
namely $C_{(1)} = D_{(0)}$.  
This determines $z_{(2)}^*$ as  
\begin{equation}
z_{(2)}^* = (z_{(1)}^*)^2 
\left(
C_{(1)} - \frac{1}{\epsilon^*} 
\right) ,  
\end{equation}
where we may use 
\begin{equation}
C_{(1)} = -z_{(1)} - \frac{z_{(1)}^*}{(\epsilon^*)^2} . 
\end{equation}

Let us consider $O(m^2)$ to look for $z_{(3)}$. 
Equation of $C_{(2)} = D_{(1)}$
provides $z_{(3)}^*$ as  
\begin{equation}
z_{(3)}^* = (z_{(1)}^*)^2 C_{(2)} 
+ \frac{(z_{(1)}^*)^3}{(\epsilon^*)^2} 
+ \frac{(z_{(2)}^*)^2}{z_{(1)}^*}  , 
\end{equation}
where we can use  
\begin{equation}
C_{(2)} = -z_{(2)} - \frac{z_{(2)}^*}{(\epsilon^*)^2} 
+ \frac{(z_{(1)}^*)^2}{(\epsilon^*)^3} . 
\end{equation}
By the same way, one can obtain 
perturbatively $n$th-order solutions $z_{(n)}$ 
around $z_{(0)} = \epsilon$. 

\begin{figure}
\includegraphics[width=84mm]{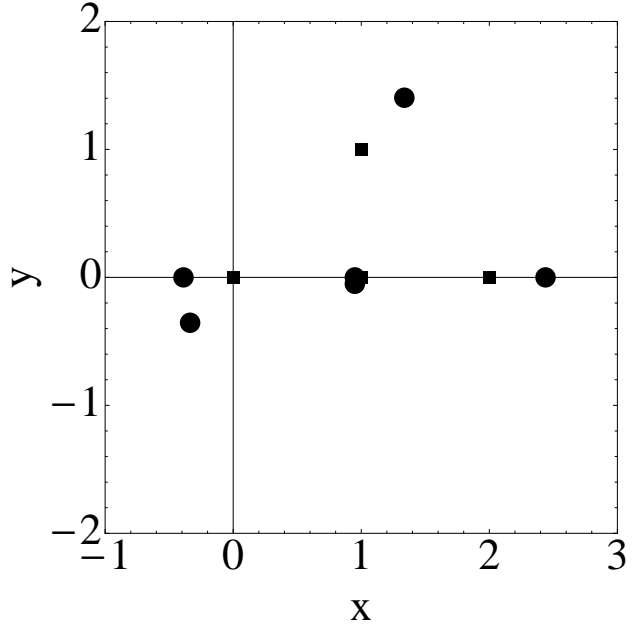}
\caption{
Perturbative image positions 
for a binary lens case. 
This plot corresponds to Tables $\ref{table-polynomial}$ 
and $\ref{table-dual}$. 
The lenses $(e_1=0, e_2=1)$ and sources $(w=2$ and $w=1+i)$ 
are denoted by filled squares. 
The image positions are denoted by filled disks. 
Perturbative images at the 1st, 2nd and 3rd orders 
are overlapped so that we cannot distinguish them 
in this figure. 
}
\label{binary}
\end{figure}

Table $\ref{table-dual}$ shows an example of perturbative roots 
in the dual-complex-variables formalism 
and their convergence.
Tables $\ref{table-polynomial}$ and $\ref{table-dual}$ 
suggest that the polynomial approach and 
the dual-complex-variables formalism are consistent 
with each other, regarding the true images. 
Figure $\ref{binary}$ shows image positions on the lens plane, 
corresponding to these tables.

\begin{table*}
\begin{minipage}{180mm}
\caption{
Example of perturbative images 
via the dual-complex-variables formalism: 
We assume the same values for parameters 
as Table $\ref{table-polynomial}$. 
Good agreements with these tables 
suggest a consistency between the single-complex-polynomial 
and the dual-complex-variables formalism, 
regarding the true images except for unphysical roots.  
}
\begin{center}
    \begin{tabular}{llll}
\hline
Case 1 (On-axis) & $\nu=0.1$ &  $e=1$ & $w=2$ 
\\
\hline
Root & 1 & 2 & 3  
\\
\hline
1st. & 2.43921 & -0.389214 & 0.95  
\\
2nd. & 2.43855 & -0.388551 & 0.95 
\\
3rd. & 2.43858 & -0.388519 & 0.949938 
\\
\hline
Lens Eq. & 2.43858 & -0.388517 & 0.949937  
\\
\hline
    \end{tabular}
  \end{center}

\begin{center}
    \begin{tabular}{llll}
\hline
Case 2 (Off-axis) & $\nu=0.1$ &  $e=1$ & $w=1+i$ 
\\
\hline
Root & 1 & 2 & 3  
\\
\hline
1st. & 1.33716+1.40546 i & -0.337158-0.355459 i & 0.95-0.05 i  
\\
2nd. & 1.33632+1.40363 i & -0.336316-0.354881 i & 0.95-0.05 i 
\\
3rd. & 1.33634+1.40371 i & -0.336275-0.354839 i & 0.95-0.05025 i  
\\
\hline
Lens Eq. & 1.33633+1.40371 i & -0.336272-0.354835 i 
& 0.950015-0.0502659 i  
\\
\hline
    \end{tabular}
  \end{center}
\label{table-dual}
\end{minipage}
\end{table*}


\section{Perturbative Solutions For $zz^*$-dual formalism 2: 
Lensing by N point mass }
The purpose of this section is to extend the proposed method 
to a general case of gravitational lensing 
by an arbitrary number of point masses. 

The lens equation is written as 
\begin{equation}
C(z, z^*) = \sum_{k=2}^N \nu_k D_k(z^*) , 
\label{C=mDk}
\end{equation}
where $C(z, z^*)$ was defined by Eq. ($\ref{C}$) 
and we defined 
\begin{eqnarray}
D_k(z^*)&=&\frac{1}{z^*}-\frac{1}{z^*-\epsilon_k^*} . 
\label{Dk}
\end{eqnarray}

$C(z, z^*)$ and $D_k(z^*)$ 
in the lens equation ($\ref{C=mDk}$) are expanded as  
\begin{eqnarray}
C(z, z^*)&=&\sum_{p_2=0}^{\infty} \sum_{p_3=0}^{\infty} 
\cdots \sum_{p_N=0}^{\infty}
(\nu_2)^{p_2} (\nu_3)^{p_3} \cdots (\nu_N)^{p_N} 
\nonumber\\
&&
\times 
C_{(p_2)(p_3) \cdots (p_N)}(z, z^*)  , 
\\
D_k(z^*)&=&\sum_{p_2=0}^{\infty} \sum_{p_3=0}^{\infty} 
\cdots \sum_{p_N=0}^{\infty}
(\nu_2)^{p_2} (\nu_3)^{p_3} \cdots (\nu_N)^{p_N} 
\nonumber\\
&&
\times
D_{k (p_2)(p_3) \cdots (p_N)}(z^*)  ,
\end{eqnarray}
where $C_{(p_2)(p_3) \cdots (p_N)}$ and 
$D_{k (p_2)(p_3) \cdots (p_N)}$ 
are independent of any $\nu_i$. 
We seek a solution in expansion series as 
\begin{equation}
z=\sum_{p_2=0}^{\infty} \sum_{p_3=0}^{\infty} \cdots
\sum_{p_N=0}^{\infty} 
(\nu_2)^{p_2} (\nu_3)^{p_3} \cdots (\nu_N)^{p_N} 
z_{(p_2)(p_3) \cdots (p_N)} , 
\label{formalsolution-dual-N}
\end{equation}
where $z_{(p_2)(p_3) \cdots (p_N)}$ is a constant 
to be determined iteratively. 
The perturbed roots are normalised by 
the zeroth-order one as 
\begin{equation}
\sigma_{(p_2)(p_3) \cdots (p_N)} 
= \frac{z_{(p_2)(p_3) \cdots (p_N)}}{z_{(0) \cdots (0)}} .
\label{N-sigma}
\end{equation}

Equation $(\ref{Dk})$ shows that 
$D_k(z^*)$ has a pole at $z^* = \epsilon_k^*$. 
Therefore, we shall discuss two cases of 
$z_{(0)} \neq \epsilon_k$ or $z_{(0)} = \epsilon_k$, separately.

\subsection{0th order 
($z_{(0) \cdots (0)} \neq \epsilon_i$ for $i=1, \cdots, N$)}
Zeroth order solutions are obtained by solving 
the equation as 
\begin{equation}
C(z, z^*) = 0.
\label{2-G0}
\end{equation} 
This has been solved for the binary lens case.  
The solution is given as 
\begin{equation}
z_{(0) \cdots (0)}=A w , 
\label{N-z0-neq}
\end{equation} 
with the coefficient $A$ defined by Eq. ($\ref{2-G3}$).

\subsection{1st order 
($z_{(0) \cdots (0)} \neq \epsilon_i$ for $i=1, \cdots, N$)} 
At the linear order in $\nu_k$, Eq. ($\ref{C=mDk}$) is 
\begin{equation}
C_{(0) \cdots (1_k) \cdots (0)} 
= \nu_k  D_{k (0) \cdots (0)} , 
\label{N-zz*1st}
\end{equation}
where $1_k$ denotes that the $k$-th index is the unity. 
This equation is rewritten as 
\begin{equation}
z_{(0) \cdots (1_k) \cdots (0)} 
+ a_{(0) \cdots (1_k) \cdots (0)} 
\times z_{(0) \cdots (1_k) \cdots (0)}^* 
= b_{(0) \cdots (1_k) \cdots (0)} ,
\label{N-zz*ab}
\end{equation}
where we defined 
\begin{eqnarray}
a_{(0) \cdots (1_k) \cdots (0)} 
&=& 
\frac{1}{(z_{(0) \cdots (0)}^*)^2} ,
\label{N-a1-neq}
\\
b_{(0) \cdots (1_k) \cdots (0)}
&=& 
\frac{\epsilon_k^*}
{z_{(0) \cdots (0)}^* (z_{(0) \cdots (0)}^* - \epsilon_k^*)} , 
\label{N-b1-neq}
\end{eqnarray}
By using Eq. ($\ref{zz*root}$), we obtain 
\begin{eqnarray}
&&z_{(0) \cdots (1_k) \cdots (0)} 
\nonumber\\
&=&
\frac{b_{(0) \cdots (1_k) \cdots (0)} - 
a_{(0) \cdots (1_k) \cdots (0)} b_{(0) \cdots (1_k) \cdots (0)}^*}
{1-a_{(0) \cdots (1_k) \cdots (0)} a_{(0) \cdots (1_k) \cdots (0)}^*} . 
\label{N-z1-neq}
\end{eqnarray}

\subsection{2nd order 
($z_{(0) \cdots (0)} \neq \epsilon_i$ for $i=1, \cdots, N$)}
Let us consider two types of second-order solutions as 
$z_{(0) \cdots (2_k) \cdots (0)}$ 
and 
$z_{(0) \cdots (1_k) \cdots (1_{\ell}) \cdots (0)}$ 
for $k \neq \ell$, separately. 

First, we shall seek $z_{(0) \cdots (2_k) \cdots (0)}$. 
At $O(\nu_k^2)$, Eq. ($\ref{C=mDk}$) becomes 
\begin{eqnarray}
&&z_{(0) \cdots (2_k) \cdots (0)} 
+ a_{(0) \cdots (2_k) \cdots (0)} z_{(0) \cdots (2_k) \cdots (0)}^* 
\nonumber\\
&&
= b_{(0) \cdots (2_k) \cdots (0)} , 
\label{zz*2ndkk}
\end{eqnarray}
where we defined 
\begin{eqnarray}
a_{(0) \cdots (2_k) \cdots (0)}&=&\frac{1}{(z_{(0) \cdots (0)}^*)^2} , 
\label{N-a2-neq}
\\
b_{(0) \cdots (2_k) \cdots (0)}&=& -D_{k(0) \cdots (1_k) \cdots (0)}
+ \frac{(\sigma_{(0) \cdots (1_k) \cdots (0)}^*)^2}{z_{(0) \cdots (0)}^*} ,
\label{N-b2-neq}
\end{eqnarray}
where $D_{k(0) \cdots (1_k) \cdots (0)}$ is written as 
\begin{eqnarray}
D_{k(0) \cdots (1_k) \cdots (0)} 
&=& 
- z_{(0) \cdots (1_k) \cdots (0)}^* 
\nonumber\\
&&
\left(
\frac{1}{(z_{(0) \cdots (0)}^{*})^2} 
- \frac{1}{(z_{(0) \cdots (0)}^* - \epsilon_k)^2}
\right) .
\label{N-D1-neq}
\end{eqnarray}

By using the relation ($\ref{zz*root}$) for Eq. ($\ref{zz*2ndkk}$), 
we obtain 
\begin{eqnarray}
&&z_{(0) \cdots (2_k) \cdots (0)} 
\nonumber\\
&=& 
\frac{b_{(0) \cdots (2_k) \cdots (0)} 
- a_{(0) \cdots (2_k) \cdots (0)} b_{(0) \cdots (2_k) \cdots (0)}^*}
{1 - a_{(0) \cdots (2_k) \cdots (0)} a_{(0) \cdots (2_k) \cdots (0)}^*} . 
\label{N-z2-neq}
\end{eqnarray}

Next, let us determine 
$z_{(0) \cdots (1_k) \cdots (1_{\ell}) \cdots (0)}$. 
At $O(\nu_k \nu_{\ell})$ for $k < \ell$, 
Eq. ($\ref{C=mDk}$) becomes 
\begin{eqnarray}
&&z_{(0) \cdots (1_k) \cdots (1_{\ell}) \cdots (0)} 
\nonumber\\ 
&& 
+ a_{(0) \cdots (1_k) \cdots (1_{\ell}) \cdots (0)}
z_{(0) \cdots (1_k) \cdots (1_{\ell}) \cdots (0)}^* 
\nonumber\\
&=& b_{(0) \cdots (1_k) \cdots (1_{\ell}) \cdots (0)} , 
\label{zz*2ndkl}
\end{eqnarray}
where we defined 
\begin{eqnarray}
a_{(0) \cdots (1_k) \cdots (1_{\ell}) \cdots (0)}
&=&\frac{1}{(z_{(0) \cdots (0)}^*)^2} , 
\label{N-a11-neq}
\\
b_{(0) \cdots (1_k) \cdots (1_{\ell}) \cdots (0)}
&=& -D_{k(0) \cdots (1_{\ell}) \cdots (0)}
-D_{\ell(0) \cdots (1_k) \cdots (0)}
\nonumber\\
&&+ \frac{2 \sigma_{(0) \cdots (1_k) \cdots (0)}^* 
\sigma_{(0) \cdots (1_{\ell}) \cdots (0)}^*}{z_{(0) \cdots (0)}^*} .
\label{N-b11-neq}
\end{eqnarray}
Here, $D_{k(0) \cdots (1_{\ell}) \cdots (0)}$ and 
$D_{\ell(0) \cdots (1_k) \cdots (0)}$
are written as 
\begin{eqnarray}
&&
D_{k(0) \cdots (1_{\ell}) \cdots (0)} 
\nonumber\\
&=& -z_{(0) \cdots (1_{\ell}) \cdots (0)}^*
\left(
\frac{1}{(z_{(0) \cdots (0)}^*)^2} 
- \frac{1}{(z_{(0) \cdots (0)}^* - \epsilon_k)^2}
\right) ,
\\
&&
D_{\ell(0) \cdots (1_k) \cdots (0)}
\nonumber\\
&=& -z_{(0) \cdots (1_k) \cdots (0)}^*
\left(
\frac{1}{(z_{(0) \cdots (0)}^*)^2} 
- \frac{1}{(z_{(0) \cdots (0)}^* - \epsilon_{\ell})^2}
\right) . 
\end{eqnarray}

By using the relation ($\ref{zz*root}$) for Eq. ($\ref{zz*2ndkl}$), 
we obtain 
\begin{eqnarray}
&&z_{(0) \cdots (1_k) \cdots (1_{\ell}) \cdots (0)}
\nonumber\\
&&=
\frac{ b_{(0) \cdots (1_k) \cdots (1_{\ell}) \cdots (0)}
- a_{(0) \cdots (1_k) \cdots (1_{\ell}) \cdots (0)} 
b_{(0) \cdots (1_k) \cdots (1_{\ell}) \cdots (0)}^*}
{1 - a_{(0) \cdots (1_k) \cdots (1_{\ell}) \cdots (0)}  
a_{(0) \cdots (1_k) \cdots (1_{\ell}) \cdots (0)}^*} . 
\nonumber\\
&&
\label{N-z11-neq}
\end{eqnarray}

\subsection{0th and 1st order ($z_{(0) \cdots (0)} = \epsilon_k$)} 
Next, we investigate the vicinity of $z = \epsilon_k$, 
which is a pole of $D_k$. 
The other pole of $D_k$ is $z=0$, which makes 
$C(z, z*)$ divergent. Therefore, $z=0$ and 
its neighbourhood are abandoned. 
Let us focus on a root around 
\begin{equation}
z_{(0) \cdots (0)} = \epsilon_k . 
\label{N-z0-eq}
\end{equation}

If we admitted $z_{(0) \cdots (1_{\ell}) \cdots (0)}$ 
around $\epsilon_k$ for $l \neq k$, 
only the $D_k$ function would contain the inverse of $\nu_{\ell}$, 
which introduces a term at $O(\nu_k/\nu_{\ell})$ in the lens equation 
and leads to inconsistency. 
Namely, the lens equation prohibits 
$z_{(0) \cdots (1_{\ell}) \cdots (0)}$ 
around $\epsilon_k$ for $l \neq k$. 
This agrees with the polynomial case. 
We thus assume $z=\epsilon_k + \nu_k z_{(0) \cdots (1_k) \cdots (0)} 
+ O(\nu_k^2)$. 
Then, we obtain 
\begin{eqnarray}
C_{(0) \cdots (0)}&=&w-\epsilon_k+\frac{1}{\epsilon_k^*} , 
\\
D_{(0) \cdots (-1_k) \cdots (0)}&=& 
- \frac{1}{z_{(0) \cdots (1_k) \cdots (0)}^*} , 
\end{eqnarray}
where $-1_k$ means that the inverse of $\nu_k$ appears 
because of the pole at $\epsilon_k$. 
Therefore, the lens equation at $O(\nu_k^0)$ becomes 
linear in $z_{(0) \cdots (1_k) \cdots (0)}^*$ 
without including $z_{(0) \cdots (1_k) \cdots (0)}$.  
Immediately, it determines $z_{(0) \cdots (1_k) \cdots (0)}^*$. 
Hence, its complex conjugate provides 
\begin{equation}
z_{(0) \cdots (1_k) \cdots (0)} 
= - \frac{\epsilon_k}{(w^* - \epsilon_k^*) \epsilon_k + 1} . 
\label{N-z1-eq}
\end{equation}

\subsection{2nd order ($z_{(0) \cdots (0)} = \epsilon_k$)} 
Here, we consider the lens equation at $O(\nu_k^1)$, 
namely $C_{(0) \cdots (1_k) \cdots (0)} = D_{(0) \cdots (0)}$, 
where 
\begin{equation}
D_{(0) \cdots (0)} = 
\frac{1}{\epsilon_k^*} 
+ \frac{z_{(0) \cdots (2_k) \cdots (0)}^*}
{(z_{(0) \cdots (1_k) \cdots (0)}^*)^2} .
\end{equation}
Hence, we obtain $z_{(0) \cdots (2_k) \cdots (0)}^*$ and 
thereby its complex conjugate as  
\begin{equation}
z_{(0) \cdots (2_k) \cdots (0)} 
= (z_{(0) \cdots (1_k) \cdots (0)})^2 
\left(
C_{(0) \cdots (1_k) \cdots (0)}^* - \frac{1}{\epsilon_k^*} 
\right) , 
\label{N-z2-eq}
\end{equation}
where $C_{(0) \cdots (1_k) \cdots (0)}$ becomes 
\begin{equation}
C_{(0) \cdots (1_k) \cdots (0)} 
= -\left(
z_{(0) \cdots (1_k) \cdots (0)} + 
\frac{z_{(0) \cdots (1_k) \cdots (0)}^*}{\epsilon_k^{*2}}
\right) . 
\end{equation}

Next, we consider a root at $O(\nu_k^1 \nu_{\ell}^1)$, 
where we can assume $k < \ell$ without loss of generality. 
At this order, the inverse of $\nu_k$ appears. 
The lens equation at $O(\nu_{\ell}^1)$ becomes 
\begin{equation}
C_{(0) \cdots (1_{\ell}) \cdots (0)} 
= D_{k (0) \cdots (-1_k) \cdots (0)} 
+ D_{\ell (0) \cdots (0)}  , 
\end{equation}
where 
\begin{equation}
D_{k (0) \cdots (-1_k) \cdots (1_{\ell}) \cdots (0)} = 
\frac{z_{(0) \cdots (1_k) \cdots (1_{\ell}) \cdots (0)}^*}
{(z_{(0) \cdots (1_k) \cdots (0)}^*)^2} . 
\end{equation}

Hence, we obtain 
$z_{(0) \cdots (1_k) \cdots (1_{\ell}) \cdots 0)}^*$ 
and thereby its complex conjugate as  
\begin{eqnarray}
&&z_{(0) \cdots (1_k) \cdots (1_{\ell}) \cdots (0)} 
\nonumber\\
&&= (z_{(0) \cdots (1_k) \cdots (0)})^2 
\nonumber\\
&&
\times
\left(
C_{(0) \cdots (1_{\ell}) \cdots (0)}^* 
- D_{\ell (0) \cdots (0)}^* 
\right) , 
\label{N-z11-eq}
\end{eqnarray}
where $C_{(0) \cdots (1_{\ell}) \cdots (0)}$ 
and $D_{\ell (0) \cdots (0)}$ are written as 
\begin{eqnarray}
C_{(0) \cdots (1_{\ell}) \cdots (0)} 
&=& - z_{(0) \cdots (1_{\ell}) \cdots (0)} 
- \frac{z_{(0) \cdots (1_{\ell}) \cdots (0)}^*}{\epsilon_k^{*2}} , 
\label{N-C1l-eq}
\\
D_{\ell (0) \cdots (0)}
&=& \frac{1}{\epsilon_k^*} 
- \frac{1}{\epsilon_k^*-\epsilon_{\ell}^*} . 
\label{N-Dl-eq}
\end{eqnarray}
This direct computation shows that 
$z_{(0) \cdots (1_k) \cdots (1_{\ell}) \cdots (0)}$ 
does not exist because 
$z_{(0) \cdots (1_{\ell}) \cdots (0)}$ 
is prohibited in the vicinity of $\epsilon_k$. 
This is also consistent with the polynomial case at the second order.


\begin{figure}
\includegraphics[width=84mm]{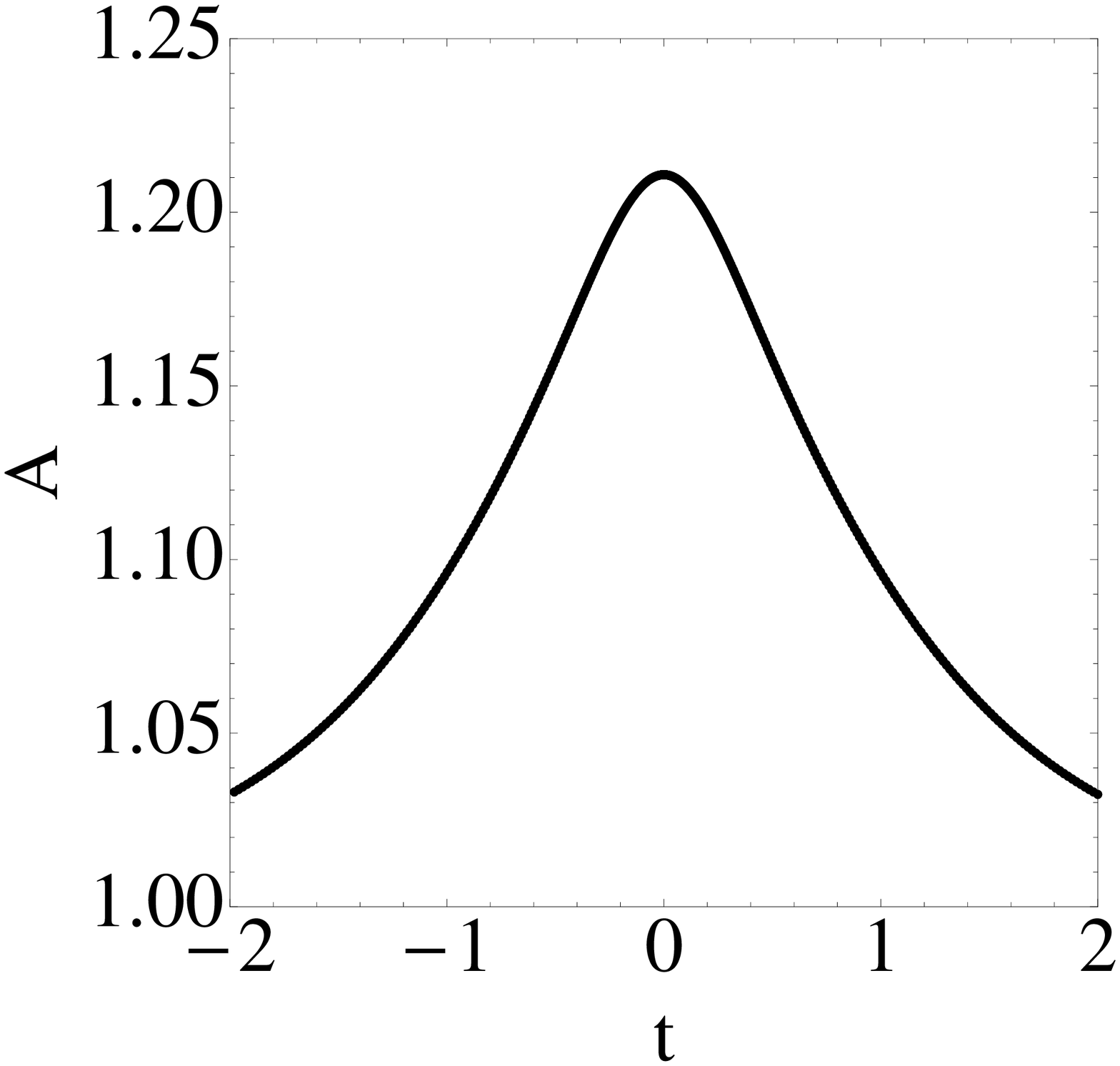}
\\
\includegraphics[width=84mm]{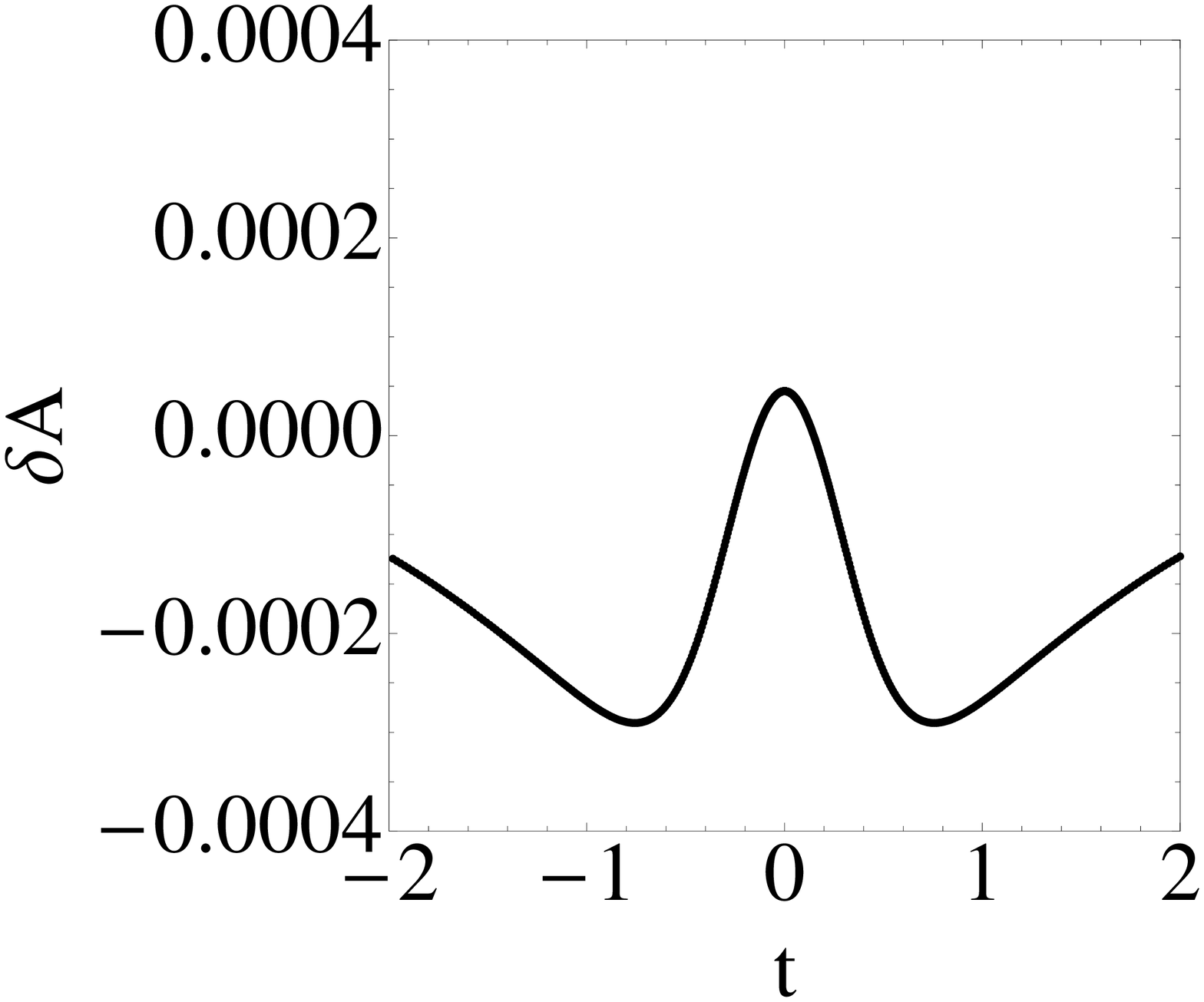}

\caption{
Light curves by two methods. 
One is based on a numerical case that 
the lens equation is solved numerically. 
The other is due to the first order approximation.  
The top figure shows that the two curves are overlapped, 
where $A$ denotes the total amplification. 
The bottom panel shows the residual by the two methods. 
The residual is defined as the difference 
between $A$ computed numerically and 
$A$ in the linear approximation. 
We assume the source trajectory as $w=1.4+i t$.
Here, the time $t$ is in units of the Einstein cross time, 
which is defined as $\theta_E/v_{\perp}$ 
for the transverse angular relative velocity. 
The lens parameters are $\nu_2=0.1$ and $e=1$. 
}
\label{lightcurve1}
\end{figure}


\begin{figure}
\includegraphics[width=84mm]{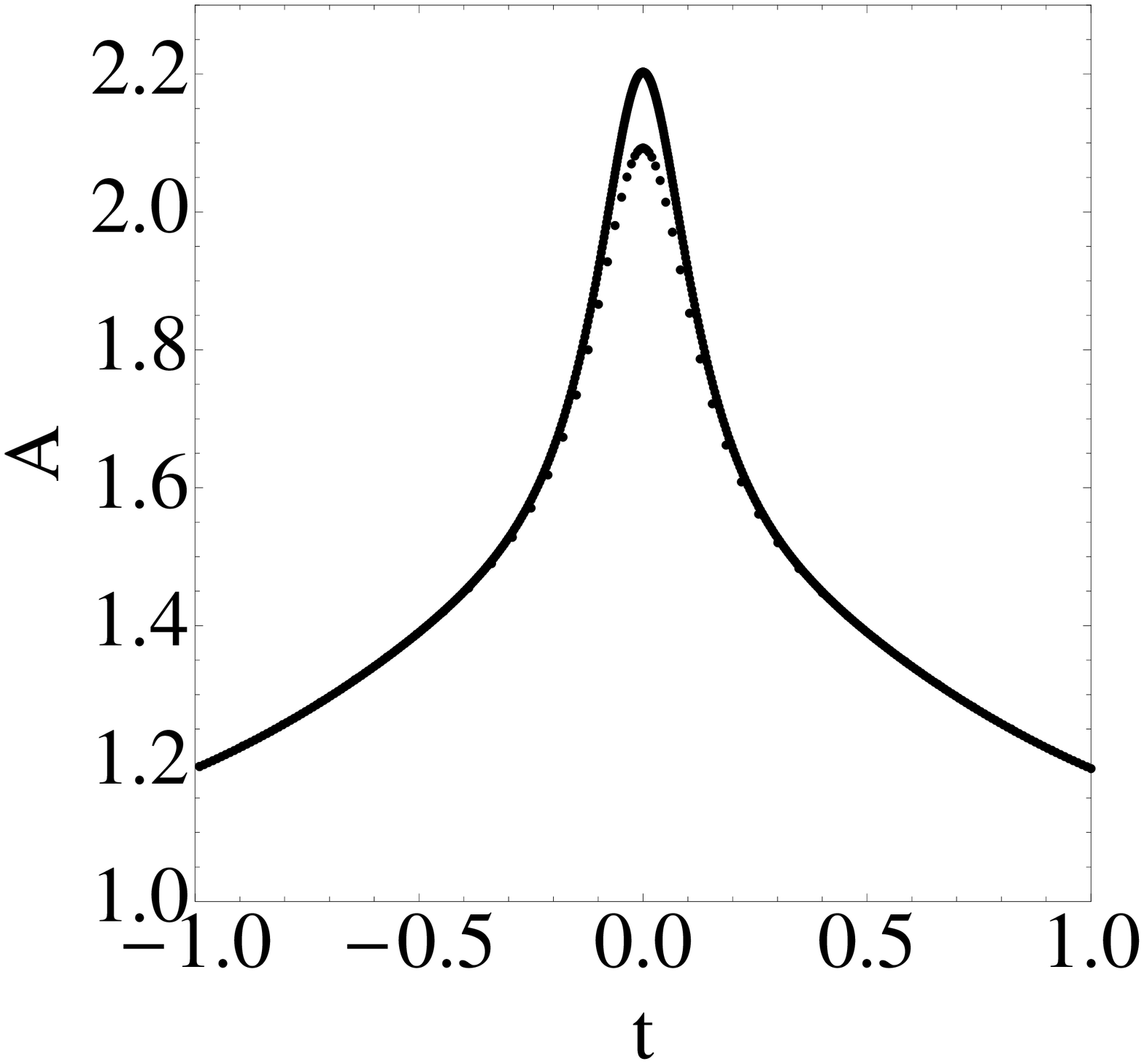}
\\
\includegraphics[width=84mm]{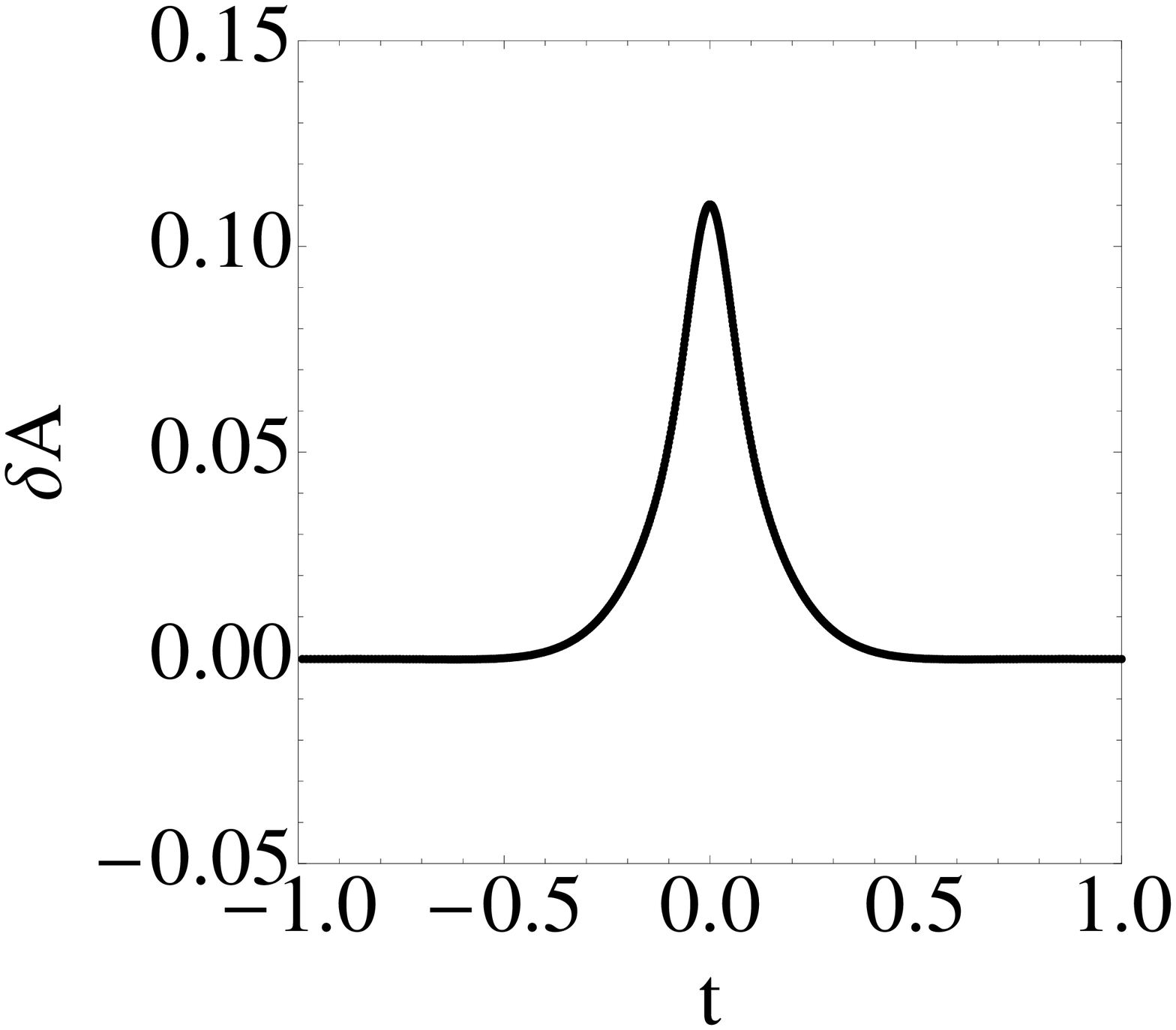}

\caption{
Light curves by two methods. 
In this figure, 
we assume a different source trajectory as $w=0.8+i t$. 
The lens parameters are the same as $\nu_2=0.1$ and $e=1$ 
in Fig. $\ref{lightcurve1}$. 
The solid curve in the top panel denotes a case 
when the lens equation is numerically solved. 
The dotted curve is drawn by using the linear order approximation.  
The bottom panel shows the residual between the two curves. 
}
\label{lightcurve2}
\end{figure}

\subsection{Magnifications} 
Before closing this paper, it is worthwhile to mention
magnifications by $N$ point-mass lensing 
in the framework of the present perturbation theory 
that is intended to solve the lens equation to obtain 
image positions. 
The amplification factor is the inverse of 
the Jacobian for the lens mapping. 
It is expressed as 
\begin{eqnarray}
A&\equiv&
\left( 
\frac{\partial\mbox{\boldmath $\beta$}}
{\partial\mbox{\boldmath $\theta$}} 
\right)^{-1}
\nonumber\\
&=&
\left( 
\frac{\partial (w, w^*)}
{\partial (z, z^*)} 
\right)^{-1}
\nonumber\\
&=&
\left( 
\left|
\frac{\partial w}{\partial z} 
\right|^2
-
\left|
\frac{\partial w}{\partial z^*} 
\right|^2
\right)^{-1} ,
\label{amplification}
\end{eqnarray}
where the terms in the last line can be computed directly 
by a derivative of Eq. ($\ref{lenseq-z}$), 
the lens equation in a complex notation. 
Amplifications of each image are obtained by substituting 
its image position into Eq. ($\ref{amplification}$). 
Practical numerical estimations may follow this procedure. 
For illustrating this, Figs. $\ref{lightcurve1}$ 
and $\ref{lightcurve2}$ show examples 
of light curves by a binary lens via the perturbative approach. 
These curves are well reproduced. 
However, double peaks due to caustic crossings cannot be 
reproduced by the present method.


\begin{figure}
\includegraphics[width=84mm]{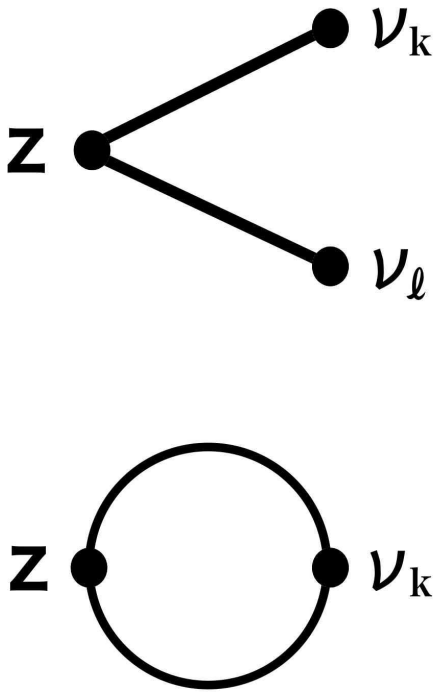}
\caption{
Graph representations of interactions among point masses 
for images at the second order level. 
The top and bottom graphs represent 
a mutually-interacting image and a self-interacting one, 
respectively.  
}
\label{graph}
\end{figure}

As an approach enabling a simpler argument 
before going to numerical estimations, 
we use the functional form of perturbed image positions. 
In the perturbation theory, lensed images can be split 
into two groups. 
One is that their zeroth-order root is not located 
at a lens object ($z_{(0) \cdots (0)} \neq \epsilon_k$). 
In the other group, zeroth-order roots originate from 
a lens position at $\epsilon_k$. 
We call the former and latter ones 
{\it mutually-interacting} and {\it self-interacting} images, respectively, 
because all the lens objects make contributions to mutually-interacting images 
at the linear order as shown by Eq. ($\ref{N-z1-neq}$). 
On the other hand, self-interacting images are influenced 
only by the nearest lens object at $\epsilon_k$ 
at the linear and even at the second orders 
as shown by Eqs. ($\ref{N-z1-eq}$) and ($\ref{N-z2-eq}$).
Figure $\ref{graph}$ shows graph representations 
for the two groups of images.

For the simplicity, we consider stretching of images 
roughly as $|\partial z/\partial w|$, 
though rigorously speaking it must be the amplification. 
Table 1 and Equation ($\ref{formalsolution-dual-N}$) mean that 
the complex derivative becomes 
for mutually-interacting images 
\begin{equation}
\frac{\partial z}{\partial w}
=
\frac{\partial z_{(0) \cdots (0)}}{\partial w}
+
\sum_k 
\nu_k 
\frac{\partial z_{(0) \cdots (1_k) \cdots (0)}}{\partial w} , 
\end{equation} 
and for self-interacting images 
\begin{equation}
\frac{\partial z}{\partial w}
=
\nu_k
\frac{\partial z_{(0) \cdots (1_k) \cdots (0)}}{\partial w} , 
\end{equation} 
where we used that $\epsilon_k$ is a constant. 

For the simplicity, we assume $\nu_k=O(1/N)$ for a large $N$ case. 
Then, the linear order term in self-interacting images is $O(1/N)$, 
and thus they become negligible as $N \to \infty$. 
On the other hand, mutually-interacting ones have non-vanishing terms 
even at the zeroth order. 
Hence, they can play a crucial role. 

\begin{figure}
\includegraphics[width=84mm]{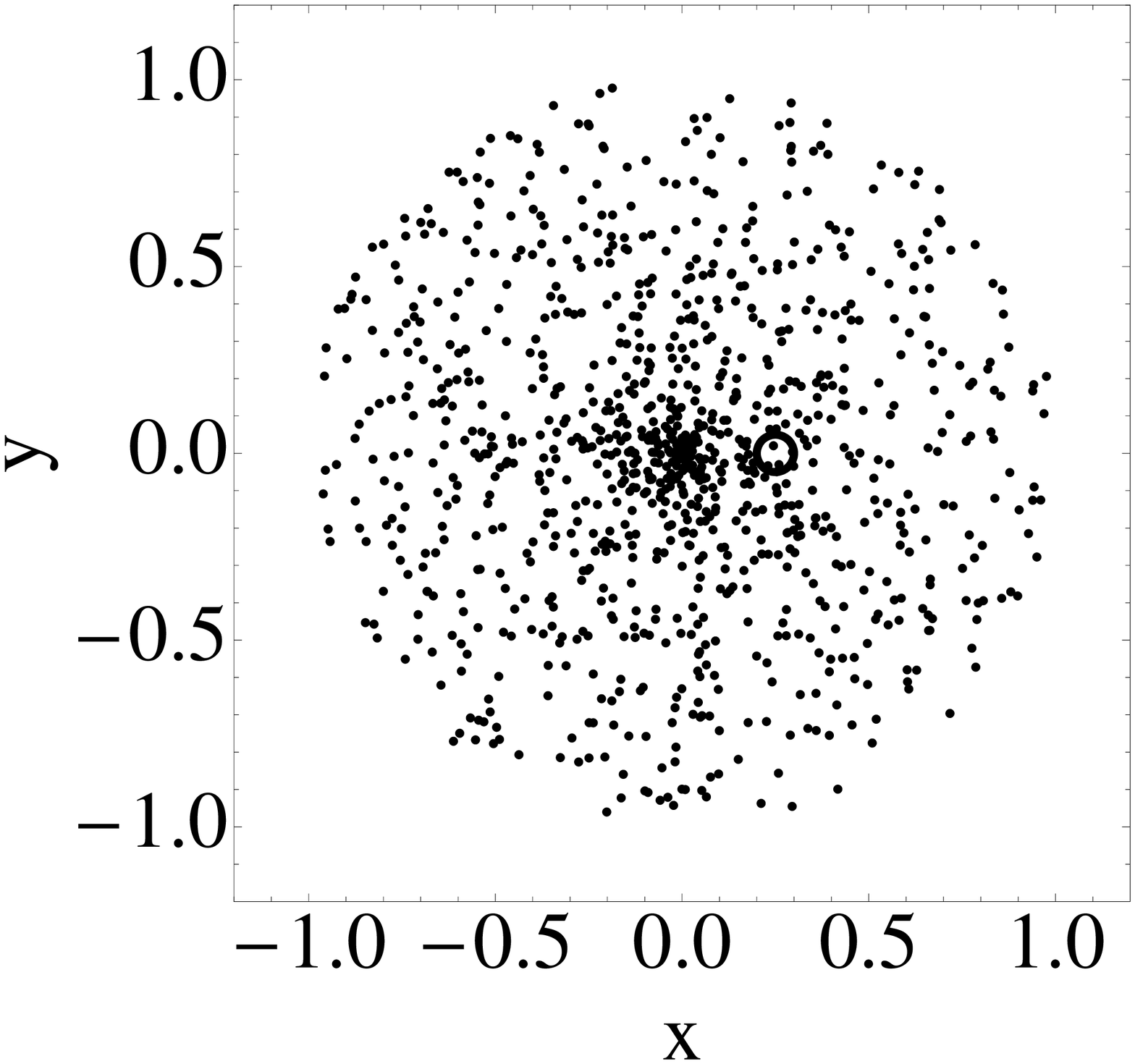}
\\
\includegraphics[width=84mm]{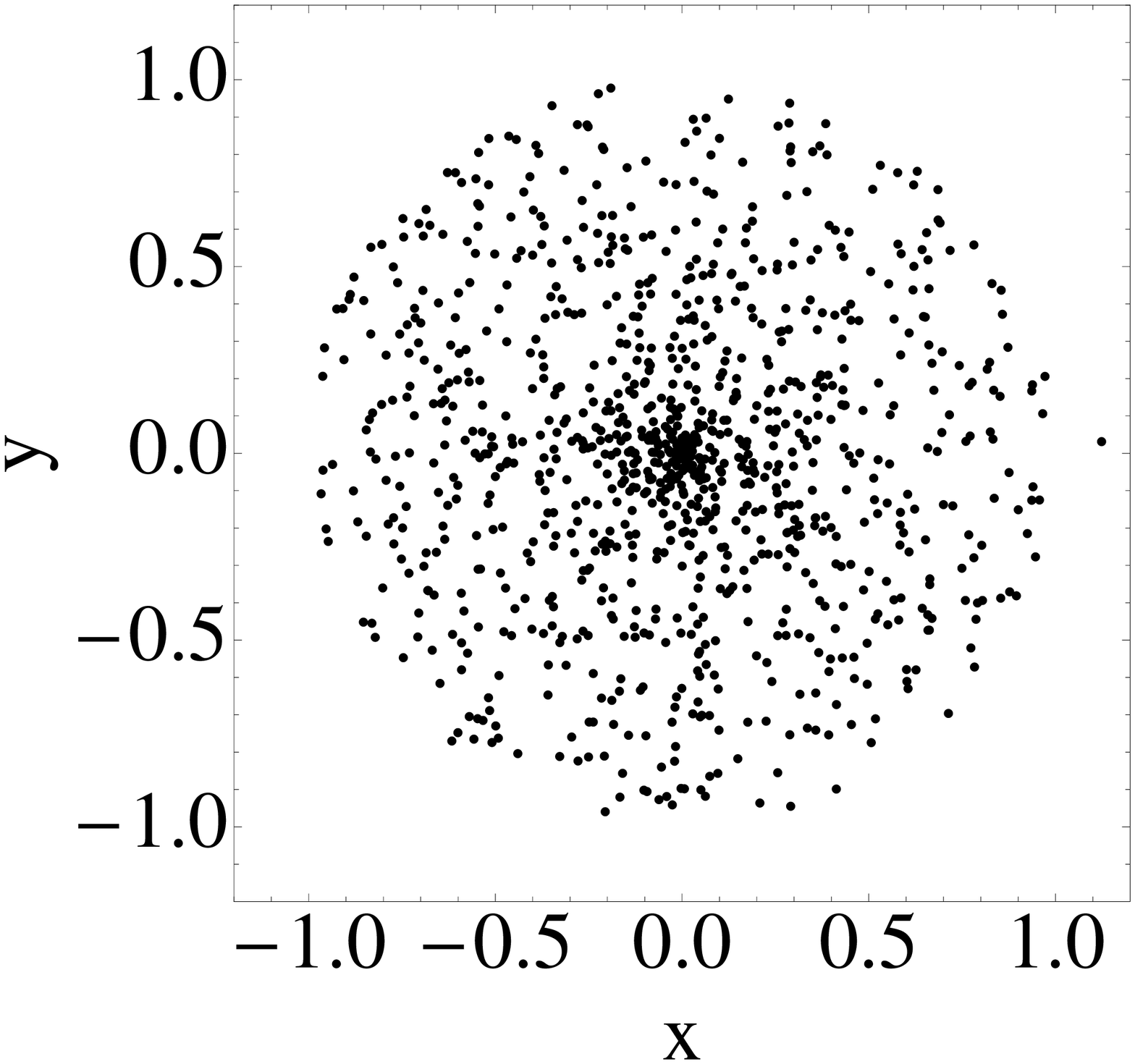}

\caption{
Example of a large N case. 
Here, we assume a truncated isothermal sphere 
projected onto a single lens plane with $N=1000$, 
where the truncation radius is the unity. 
For the simplicity, we assume equal masses. 
The source located at 0.25 is denoted by the circle. 
The top figure shows locations of 
the N point masses on the lens plane. 
The bottom shows a plot of image positions by using 
the perturbative solutions at the second order. 
In practice, the linear-order and second-order roots 
make no difference distinguishable by eyes in the figure. 
}
\label{demo-N}
\end{figure}

\begin{figure}
\includegraphics[width=84mm]{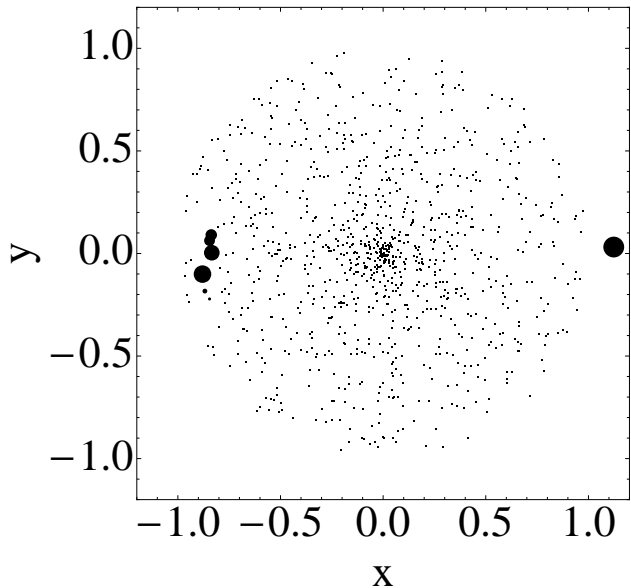}
\caption{
Plot of image positions with lensing amplification 
for a case of $N=1000$. 
The source and lenses are the same as those in Fig. $\ref{demo-N}$. 
Here, we take account of amplifications by lensing. 
The area of a disk corresponding to each image 
is proportional to the magnification factor 
in arbitrary units. 
Large amplifications near $\pm1$ are caused by 
the mutually-interacting images. 
On the other hand, a concentration of small but many images 
around the center are due to the self-interacting images, 
because lens objects have a large number density there. 
These three regions may correspond to three images 
for a singular isosphere lens in the limit of $N \to \infty$. 
}
\label{demo-N2}
\end{figure}

However, we should take account of a spatial distribution 
of lens objects. If they are clustering and thus 
dense at a certain region, then the total flux of light 
through such a dense region is not negligible any more. 
Let us denote the fraction of the clustering particles 
by $f$. 
Total contributions from such clustering self-interacting images 
are estimated approximately as a typical image magnification 
multiplied by the number of the particles,  
namely  
$f N \times \nu_k(\sim 1/N) = O(f)$, 
which does not vanish even as $N \to \infty$. 
Figures $\ref{demo-N}$ and $\ref{demo-N2}$ show 
an example of a large $N$ case, 
where $N$ is chosen as 1000. 



\begin{figure}
\includegraphics[width=84mm]{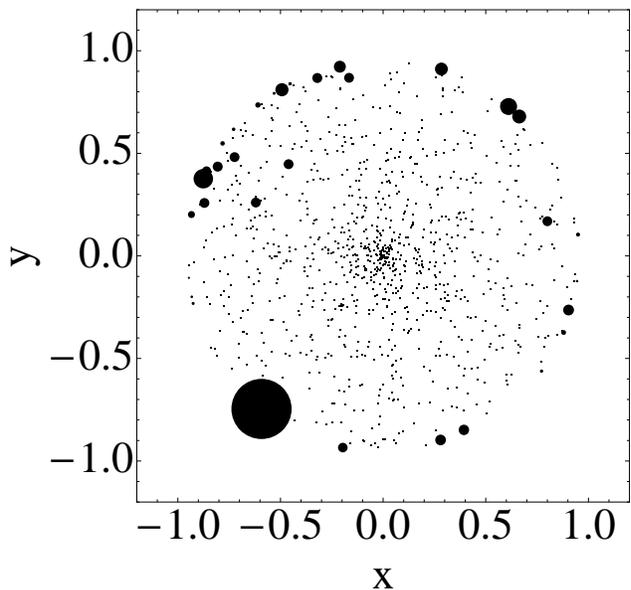}
\caption{
Einstein ring broken by the lens discreteness due to 
the finite-$N$ effect. 
The lenses are the same as those in 
Figs. $\ref{demo-N}$ and $\ref{demo-N2}$. 
The source is located at the origin of the coordinates. 
Amplifications are taken into account. 
The area denotes the magnification factor 
in arbitrary units. 
}
\label{demo-N-ring}
\end{figure}

\section{Conclusion} 
Under a small mass-ratio approximation, 
this paper developed a perturbation theory 
of N coplanar (in the thin lens approximation) 
point-mass gravitational lens systems 
without symmetries on a plane. 
The system can be separated into a single mass lens 
as a background and its perturbation 
due to the remaining point masses. 

First, we investigated perturbative structures of 
the single-complex-variable polynomial, 
into which the lens equation is embedded. 
Some of zeroth-order roots of the polynomial 
do not satisfy the lens equation and thus are unphysical. 
This appearance of correct but unphysical roots 
is consistent with the earlier work on a theorem  
on the maximum number of lensed images (Rhie 2001, 2003). 
However, the theorem  never tells which 
roots are physical (or unphysical).
What we did is that unphysical roots are identified. 

Next, we re-examined the lens equation 
in the dual-complex-variables formalism 
to avoid inclusions of unphysical roots. 
We presented an explicit form of perturbed image positions 
as a function of source and lens positions. 
As a key tool for perturbative computations, 
Eq. ($\ref{zz*root}$) was also found. 
For readers' convenience, the perturbative roots 
are listed in Table $\ref{list}$. 
If one wishes to go to higher orders, 
our method will enable one to easily use 
computer algebra softwares such as MAPLE and MATHEMATICA. 
This is because it requires simpler algebra 
(only the four basic operations of arithmetic), 
compared with vector forms which need extra operations 
such as inner and outer products. 

There are numerous possible applications 
along the course of the perturbation theory 
of N point-mass gravitational lens systems.  
For instance, it will be interesting to study 
lensing properties such as magnifications 
by using the functional form of image positions. 
Furthermore, the validity of the present result 
may be limited in the weak field regions. 
It is important also to extend the perturbation theory 
to a domain near the strong field.  

Our method considers only the images which exist 
in the small mass limit as $\nu_i \to 0$. 
The number of the images that admit the small mass-ratio limit 
is less than the maximum number. 
This suggests that the other images do not have the small mass limit. 
Therefore, it is conjectured that positions of the extra images 
could not be expressed as Maclaurin series in mass ratios. 
This may be implied also by previous works. 
For instance, the appearance of the maximum number of images 
for a binary lens requires a finite mass ratio and the caustic crossing 
(Schneider and Weiss 1986).  
Regarding this point, further studies will be needed to 
determine positions of all the images with the maximum number 
as a function of lens and source parameters.

\section*{Acknowledgments}
The author would like to thank S. Mao, N. Rattenbury and E. Kerins 
for the hospitality at the Manchester Microlensing Conference, 
where this work was initiated. 
He is grateful also to D. Bennett and Y. Muraki 
for stimulating conversations at the conference. 
He wishes to thank F. Abe, J. Bayer and D. Khavinson for 
useful comments on the manuscript. 
This work was supported in part by a Japanese Grant-in-Aid 
for Scientific Research from the Ministry of Education, 
No. 19035002.
 

\begin{table}
\caption{
List of the coefficients in perturbative positions 
of images lensed by N point masses: 
The image positions are expressed in the form of 
$
z=\sum_{p_2} \cdots
\sum_{p_N} 
(\nu_2)^{p_2} \cdots (\nu_N)^{p_N} 
z_{(p_2) \cdots (p_N)} . 
$ 
The top and bottom panels show 
the cases of $z_{(0) \cdots (0)} = \epsilon_i$ 
and $z_{(0) \cdots (0)} \neq \epsilon_i$, respectively. 
In the columns, ''None'' means that 
the corresponding coefficient does not exist. 
}
\label{table1}  
\begin{center}
    \begin{tabular}{ll}
Case 1: & $z_{(0) \cdots (0)} \neq \epsilon_i \: (i=1, \cdots, N)$ 
\\
\hline
$z_{(0) \cdots (0)}$ & Eq. ($\ref{N-z0-neq}$) 
\\
\hline
$z_{(0) \cdots (1_k) \cdots (0)}$ 
& Eqs. ($\ref{N-a1-neq}$)-($\ref{N-z1-neq}$)
\\
\hline
$z_{(0) \cdots (2_k) \cdots (0)}$ 
& Eqs. ($\ref{N-a2-neq}$)-($\ref{N-z2-neq}$) 
\\
\hline
$z_{(0) \cdots (1_k) \cdots (1_{\ell}) \cdots (0)}$ 
& Eqs. ($\ref{N-a11-neq}$)-($\ref{N-z11-neq}$)
    \end{tabular}
  \end{center}

\label{table2}  
\begin{center}
    \begin{tabular}{ll}
Case 2: & $z_{(0) \cdots (0)} = \epsilon_k$ 
\\
\hline
$z_{(0) \cdots (0)}$ & Eq. ($\ref{N-z0-eq}$) 
\\
\hline
$z_{(0) \cdots (1_k) \cdots (0)}$ 
& Eqs. ($\ref{N-z1-eq}$)
\\
\hline
$z_{(0) \cdots (1_{\ell}) \cdots (0)}$ for $\ell \neq k$ 
& None
\\
\hline
$z_{(0) \cdots (2_k) \cdots (0)}$ 
& Eq. ($\ref{N-z2-eq}$) 
\\
\hline
$z_{(0) \cdots (1_k) \cdots (1_{\ell}) \cdots (0)}$ for $\ell \neq k$  
& None 
    \end{tabular}
  \end{center}

\label{list}
\end{table}

\bsp

\label{lastpage}

\end{document}